\documentclass[12pt]{iopart}

\pdfoutput=1

\usepackage{graphicx,epsfig,sidecap}
\usepackage{bm}
\usepackage{braket}
\usepackage{amsfonts,amssymb}

\usepackage{color}
\usepackage{times}
\usepackage[colorlinks,bookmarks=false,citecolor=blue,linkcolor=red,urlcolor=blue]{hyperref}

\def\be{\begin{equation}}
\def\ee{\end{equation}}

\def\bea{\begin{eqnarray}}
\def\eea{\end{eqnarray}}

\def\Tr{{\rm Tr}}

\def\e{\epsilon}
\def\s{\sigma}

\def\l{\lambda}
\def\HH{\mathcal H}

\begin{document}

\title[Entanglement negativity in the critical Ising chain]
{Entanglement negativity in the critical Ising chain}

\author{Pasquale Calabrese$^1$, Luca Tagliacozzo$^2$, and Erik Tonni$^3$}
\address{$^1$ Dipartimento di Fisica dell'Universit\`a di Pisa and INFN,
             Pisa, Italy.\\
          $^2$ ICFO - The Institute of Photonic Sciences, Av. C.F. Gauss 3, E-08860 Castelldefels (Barcelona), Spain.\\
          $^3$ SISSA and INFN, via Bonomea 265, 34136 Trieste, Italy. }

\date{\today}

\begin{abstract}

We study the scaling of the traces of the integer powers of the partially transposed reduced density matrix 
$\Tr (\rho_A^{T_2})^n$ and of the entanglement negativity for two spin blocks as function of their length and separation 
in the critical Ising chain.
For two adjacent blocks, we show that tensor network calculations agree
with universal conformal field theory (CFT) predictions. 
In the case of two disjoint blocks the CFT predictions are recovered  
only after taking into account the finite size corrections induced by the finite length 
of the blocks.

\end{abstract}

\maketitle

\section{Introduction}

During the last decade, it became clear that new concepts from quantum information theory can be 
extremely useful also to characterize many-body quantum systems both in and out of equilibrium. 
The most successful of these concepts is surely the entanglement entropy  
which turned out to be an optimal probe to distinguish the various states of 
matter, in particular for critical and topological phases (see e.g. \cite{rev} for reviews). 
%
Let $\rho$ be the density matrix of an extended quantum system, which we take to be
in a pure quantum state $|\Psi\rangle$, so that $\rho=|\Psi\rangle\langle\Psi|$. 
Let the Hilbert space be written as a direct product $\HH=\HH_A\otimes\HH_B$. $A$'s reduced density
matrix is $\rho_A=\Tr_B \rho$ and from this the R\'enyi entanglement entropy is defined as
\be 
S^{(n)}_A= \frac{1}{1-n} \ln {\rm Tr}\,\rho_A^n\,, 
\label{Sndef}
\ee
that for $n=1$ reduces to the more studied von Neumann entropy. 
However, the knowledge of the R\'enyi entropies for any $n$ provides more 
information than the $n=1$ case and, in fact, it gives  
the full spectrum of the reduced density matrix \cite{cl-08}.

For a one-dimensional critical system whose scaling limit is described by a conformal
field theory (CFT), in the case when A is an interval of length $\ell$ 
embedded in an infinite system, the asymptotic large $\ell$ behavior of the R\'enyi
entropies is  \cite{Holzhey,cc-04,Vidal,cc-rev}
\begin{equation}
\label{Renyi:asymp}
S_A^{(n)}=\frac{c}6 \left(1+\frac1n\right)\log \frac{\ell}a +c'_n\,,
\end{equation}
where $c$ is the central charge of the underlying CFT  \cite{c-lec} 
and $a$ the inverse of an ultraviolet cutoff (e.g. the lattice spacing).
The  additive constants $c'_n$ are non universal 
but satisfy some universal relations \cite{fcm-10}. 

Unfortunately, the entanglement entropies do not provide any information about the 
multipartite entanglement in a many-body system. 
In order to give a practical example, let us imagine to divide an extended quantum system in three spatial parts 
which we call $A_1$, $A_2$ and $B$. 
We can  define the reduced density matrix  
$\rho_{A_1\cup A_2}=\Tr_B \rho$, but the corresponding entropies $S^{(n)}_{A_1\cup A_2}$ would measure 
only the entanglement between $A=A_1\cup A_2$ and $B$, but not the entanglement 
{\it between} $A_1$ and $A_2$ which is the tripartite entanglement in the system.
Even the R\'enyi mutual information $I_{A_1:A_2}^{(n)}=S_{A_1}^{(n)}+S_{A_2}^{(n)}-S_{A_1\cup A_2}^{(n)}$ does not provide the  
entanglement between $A_1$ and $A_2$, but gives only a measure of their correlations (see e.g. Ref. \cite{pv-07}). 
The quantification of the tripartite entanglement in a pure state has been longly a problem
(see e.g. Refs. \cite{rev,pv-07,v-98,ep-99,varmix}) until a   computable measurement of entanglement
has been introduced by Vidal and Werner \cite{vw-01}. 
This measure has been called negativity and  is defined as follows.
Let us denote by $|e_i^{(1)}\rangle$ and $|e_j^{(2)}\rangle$ two  bases 
in the Hilbert spaces  corresponding to $A_1$ and $A_2$ respectively. 
Let us define the partial transpose  with respect to $A_2$ degrees of freedom  as
\be 
\langle e_i^{(1)} e_j^{(2)}|\rho^{T_2}_{A_1\cup A_2}|e_k^{(1)} e_l^{(2)}\rangle=
\langle e_i^{(1)} e_l^{(2)}|\rho_{A_1\cup A_2}| e^{(1)}_k e^{(2)}_j\rangle,
\label{rhoAT2def}
\ee
and from this the  logarithmic negativity as
\be
{\cal E}\equiv\ln ||\rho^{T_2}_{A_1\cup A_2}||=\ln \Tr |\rho^{T_2}_{A_1\cup A_2}|\,,
\label{negdef}
\ee
where the trace norm  $||\rho^{T_2}_{A_1\cup A_2}||$ is
the sum of the absolute values of the eigenvalues $\lambda_i$ of $\rho^{T_2}_{A_1\cup A_2}$.
The negativity has been used to investigate the tripartite entanglement content of many 
body quantum systems both in their ground-state 
\cite{Audenaert02,Neg1,Neg2,Neg3,sod2,kor1,kor2,sod1} or out of equilibrium \cite{sod4,sod3}.
Furthermore, the definition of the negativity  is very intriguing because it is basis independent and so calculable 
by quantum field theory (QFT). 
Only recently \cite{us-letter,us-long} a systematic and generic method  to calculate the negativity  in
QFT (and in particular  CFT) has been developed. 
This method is  based on the calculation of the traces $\Tr (\rho_A^{T_2})^n$
and the negativity is recovered in a replica limit. 
The negativity between two adjacent intervals has been calculated for a general CFT 
and  turned out to be a universal quantity depending  only on the central charge. 
The negativity between two disjoint intervals is still universal but depends on the full operator content of the theory and 
the corresponding moments  $\Tr (\rho_{A_1\cup A_2}^{T_2})^n$ have been calculated explicitly 
for the free compactified boson \cite{us-long}.
In this manuscript we extend the previous results to the critical Ising CFT
and confirm our predictions by explicit tensor network calculations in the transverse field Ising chain
described by the Hamiltonian
\be
H=-\sum_{j=1}^L [\s_j^x \s_{j+1}^x+h \s_j^z]\,,
\label{IsingH def}
\ee
where $\s_j^{x,z}$ are Pauli matrices acting on the spin at site $j$ and we use
periodic boundary conditions.
The model has a quantum critical point  at $h=1$ and 
in the continuum limit corresponds to a free massless Majorana fermion which is a CFT 
with central charge $c=1/2$. 

The manuscript is organized as follows. 
In Sec. \ref{Sec2} we discuss the general QFT approach to negativity \cite{us-letter,us-long}
and report those results which are valid for an arbitrary CFT.
In Sec. \ref{Sec3} we report the CFT calculation of the integer powers of the partial transpose of the 
reduced density matrix for two disjoint intervals in the critical Ising theory.
Although the Ising chain is mapped to free fermions, it is still not known how to calculate effectively 
the partial transpose and for this reason we resort purely numerical methods. 
In Sec. \ref{Sec4} we introduce the tree tensor network approach and explain how to adapt it to the 
calculation of the eigenvalues of the partially transposed reduced density matrix. 
Finally we analyze the numerical data in  Sec. \ref{Sec5} taking properly into account the 
corrections to the scaling.
In Sec. \ref{Sec6} we draw our conclusions.

\section{General results for the negativity in CFT: one interval and two adjacent intervals}
\label{Sec2}

In the following we will be interested in tripartion of a one-dimensional systems such as those
depicted in Fig. \ref{intervals}.
We will denote with $\rho_A$ the reduced density matrix of $A\equiv A_1\cup A_2$, 
i.e. $\rho_A=\rho_{A_1\cup A_2}$, which is obtained by tracing out the part $B$ of the system, i.e. $\rho_A=\Tr_B \rho$.  
The quantum field theory approach of  negativity  is based on a replica trick \cite{us-letter,us-long}, i.e.  
on the calculation of the traces $\Tr (\rho_A^{T_2})^n$ of integer powers of $\rho_A^{T_2}$. 
According to Eq. (\ref{negdef})
we are interested in the sum of the absolute values of the eigenvalues $\lambda_i$ of $\rho_A^{T_2}$.
It turns out that $\Tr (\rho_A^{T_2})^n$ have a different functional dependence on $|\lambda_i|$ 
according to the parity of $n$.
Indeed,  for $n$ even and odd (that we denote as $n_e$ and $n_o$ respectively), 
the traces of integer powers of $\rho_A^{T_2}$ are
\bea
\Tr (\rho_A^{T_2})^{n_e}&=&\sum_i \lambda_i^{n_e}= \sum_{\l_i>0} |\l_i|^{n_e}+ \sum_{\l_i<0} |\l_i|^{n_e}\,, 
\label{trne}\\
\Tr (\rho_A^{T_2})^{n_o}&=&\sum_i \lambda_i^{n_o}= \sum_{\l_i>0} |\l_i|^{n_o}- \sum_{\l_i<0} |\l_i|^{n_o}\,.
\label{trno}
\eea
If we set $n_e=1$ in Eq. (\ref{trne}) we formally obtain $ \Tr |\rho_A^{T_2}|$ in which we are 
interested. Instead, setting $n_o=1$ in Eq. (\ref{trno}) gives the normalization $\Tr \rho_A^{T_2}=1$.
This means that the analytic continuations from even and odd $n$ are different and 
 the trace norm in which we are interested is obtained
by considering the analytic continuation of the even sequence at $n_e\to1$, 
i.e. 
\be  {\cal E}=\lim_{n_e\to1} \ln \Tr (\rho_A^{T_2})^{n_e}\,.
\ee

In a QFT, the traces of integer powers of the partial transpose for two disjoint intervals  
(as in Fig. \ref{intervals} (top))
are partition functions on $n$-sheeted Riemann surfaces or equivalently 
the correlation functions of four twist-fields \cite{us-long}
\be
\Tr(\rho_A^{T_{2}})^n=
\langle {\cal T}_n(u_1)\overline{\cal T}_n(v_1) \overline{\cal T}_n(u_2){\cal T}_n(v_2)\rangle\,,
\label{4ptdef}
\ee
i.e. the partial transposition has the net effect to exchange two twist operators compared to 
\be
\Tr\rho_A^n=\langle {\cal T}_n(u_1)\overline{\cal T}_n(v_1) {\cal T}_n(u_2)\overline{\cal T}_n(v_2)\rangle\,.
\label{rhoatw}
\ee
(See Refs. \cite{ccd-09,cc-rev} for an introduction to the concept of twist fields.)

Eq. (\ref{4ptdef}) is of general validity, but it simplifies when 
specialized to the case of two adjacent intervals, obtained by letting $v_1\to u_2$,  giving
the three-point function
\be
\Tr(\rho_A^{T_{2}})^n=
\langle {\cal T}_n(u_1) \overline{\cal T}_n^2(u_2){\cal T}_n(v_2)\rangle\,.
\label{3ptdef}
\ee
A further simplification occurs when specializing to a pure state by letting $B\to \emptyset$ 
(i.e. $u_2\to v_1$ and $v_2\to u_1$) for which 
$\Tr (\rho_A^{T_2})^{n}$ becomes a two-point function 
\be
\Tr (\rho_A^{T_2})^{n}=\langle {\cal T}^2_{n}(u_2) \overline{\cal T}^2_{n}(v_2)\rangle\,.
\ee
As explained in more details in Ref.~\cite{us-long}, as a partition function on an $n$-sheeted 
Riemann surface, this expression depends on the parity of $n$ because  
${\cal T}_n^2$ connects the $j$-th sheet with the $(j+2)$-th one. 
For $n=n_e$ even, 
the $n_e$-sheeted Riemann surface decouples in two independent ($n_e/2$)-sheeted 
surfaces.
Conversely for $n=n_o$ odd, the surface remains a $n_o$-sheeted Riemann
surface. In formulas, these observations are
\bea
\Tr (\rho_A^{T_2})^{n_e}&=& (\langle {\cal T}_{n_e/2}(u_2) \overline{\cal T}_{n_e/2}(v_2)\rangle)^2=
(\Tr\rho_{A_2}^{n_e/2})^2\,,
\label{pureqfte} \\
\Tr (\rho_A^{T_2})^{n_o}&= & \langle {\cal T}_{n_o}(u_2) \overline{\cal T}_{n_o}(v_2)\rangle=\Tr \rho_{A_2}^{n_o}\,.
\label{pureqft}
\eea
Hence, for a bipartite system, $\Tr (\rho_A^{T_2})^{n}$ can be generically written as a function of 
$\Tr \rho_{A_2}^n$, as it should. 
In particular, taking the limit $n_e\to1$, we obtain the logarithmic negativity ${\cal E}= S_{A_2}^{(1/2)}$
which is the well known result that for bipartite states the logarithmic negativity equals the 
R\'enyi entropy of order $1/2$ \cite{vw-01}.

\subsection{A single interval in a CFT.}

For conformal invariant theories it is useful to consider first  the case of a reduced 
density matrix corresponding to a bipartite system obtained by letting $B\to \emptyset$.
We recall first the standard result for $\Tr \rho_{A_2}^n$ in a bipartite pure state 
in the case when $A_2$ is an interval of length $\ell=|u_2-v_2|$ in the infinite line \cite{cc-04}
\be
\Tr \rho_{A_2}^n= \langle {\cal T}_{n}(u_2) \overline{\cal T}_{n}(v_2)\rangle = c_n\left(\frac\ell{a} \right)^{-c/6(n-1/n)}\,,
\label{trnasy}
\ee
i.e. the twist fields behave like primary operators with scaling dimension
$\Delta_{{\cal T}_n}=\Delta_{ \overline{\cal T}_{n}}=c/12(n-1/n)$.

Following Refs. \cite{us-letter,us-long}, when the interval $A_2$ is embedded in an infinite system, 
we can derive the powers of $\rho^{T_2}_A$ using Eqs. (\ref{pureqfte}) and (\ref{pureqft})
and from the general formula for $\Tr \rho_{A_2}^n$ in Eq. (\ref{trnasy}), obtaining 
\be\fl
\Tr (\rho_A^{T_2})^{n_e}= (\langle {\cal T}_{n_e/2}(u_2) \overline{\cal T}_{n_e/2}(v_2)\rangle)^2=(\Tr\rho_{A_2}^{n_e/2})^2
=c_{n_e/2}^2 \Big(\frac{\ell}a\Big)^{-{c}/{3}({n_e}/2-2/{n_e})}\,,
\label{1inte}
\ee
and 
\be\fl
\Tr (\rho_A^{T_2})^{n_o}=  \langle {\cal T}_{n_o}(u_2) \overline{\cal T}_{n_o}(v_2)\rangle=\Tr \rho_{A_2}^{n_o}=
c_{n_o} \Big(\frac{\ell}a\Big)^{-{c}/6(n_o-1/{n_o})}.
\label{1into}
\ee
This simple result shows an important feature of the negativity in CFT, i.e.
for $n=n_e$ even,  
${\cal T}^2_{n_e}$ and $\overline{\cal T}^2_{n_e}$ have dimensions 
\be
\Delta_{{\cal T}_{n_e}^2}=\Delta_{\overline{\cal T}_{n_e}^2}=\frac{c}6 \Big(\frac{n_e}2-\frac2{n_e}\Big), 
\label{Dtne}
\ee
while  for $n=n_o$ odd,  ${\cal T}^2_{n_o}$  and $\overline{\cal T}^2_{n_o}$ have dimensions 
\be 
\Delta_{{\cal T}_{n_o}^2}=\Delta_{\overline{\cal T}_{n_o}^2}=\frac{c}{12}\Big(n_o-\frac1{n_o}\Big),
\label{Dtno}
\ee
the same as ${\cal T}_{n_o}$.  
Thus, performing the analytic continuation from the even sequence, we finally have 
\be\fl
|| \rho_A^{T_2}||= \lim_{n_e\to1} \Tr (\rho_A^{T_2})^{n_e}=
c_{1/2}^2  \Big(\frac{\ell}a\Big)^{{c}/2}\qquad  \Rightarrow\;\; {\cal E}=\frac{c}2\ln \frac{\ell}a+2\ln c_{1/2}\,,
\label{neg2pt}
\ee
which again is the result that for pure bipartite states the logarithmic negativity equals the 
R\'enyi entropy of order $1/2$.
Continuing instead the odd sequence to $n_o\to1$,  we recover the normalization $\Tr\rho_A^{T_2}=1$.

Notice that  the constants $c_{n_o,n_e}$ are non-universal, but they are the same 
as in the entanglement entropies and so new non-universal constants do not appear
in the partial transposition, but are all already encoded in the reduced density matrix.

\subsection{Two adjacent intervals in a CFT.}

\begin{figure}[t]
\begin{center}
\includegraphics[width=.9\textwidth]{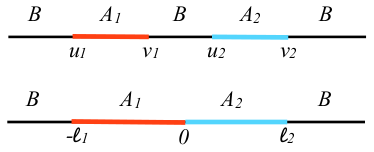}
\end{center}
\caption{The two configurations of the Ising chain that  we consider.
Top: the entanglement between two disjoint intervals $A_1$ and $A_2$ embedded 
in the ground-state of a larger chain.
Bottom: The entanglement between two adjacent intervals in a larger system.
}
\label{intervals}
\end{figure}

Let us now consider  the configuration of two intervals $A_1$ and $A_2$ of length $\ell_1$ and $\ell_2$ 
sharing a common boundary  at the origin as graphically depicted in Fig.~\ref{intervals} (bottom).
This can be obtained by letting $v_1\to u_2=0$ in Eq. (\ref{4ptdef}) and it is then described by the  3-point function
(we set $u_1=-\ell_1$ and $v_2=\ell_2$)
\be
\Tr (\rho_A^{T_{2}})^{n}=\langle {\cal T}_{n}(-\ell_1) \overline{\cal T}^2_{n}(0){\cal T}_{n}(\ell_2) \rangle\,,
\ee
whose form is determined by global conformal symmetry \cite{cft-book}
\be
\langle {\cal T}_{n}(-\ell_1) \overline{\cal T}^2_{n}(0){\cal T}_{n}(\ell_2) \rangle=
 \frac{d_n}{
(\ell_1\ell_2)^{\Delta_{{\cal T}_n^2}} (\ell_1+\ell_2)^{2\Delta_{{\cal T}_n}-\Delta_{{\cal T}_n^2}}},
\label{3ptgen}
\ee
where, for simplicity, we have set the UV cutoff $a=1$.
Notice that $d_n$ is not universal, but can be written as 
$d_n=c_n^2 C_{{\cal T}_{n} \overline{\cal T}^2_{n} {\cal T}_{n}}$
where the structure constant $C_{{\cal T}_{n} \overline{\cal T}^2_{n} {\cal T}_{n}}$ is universal 
and can be determined by considering the proper limit of the four-point function as explicitly done in Ref. \cite{us-long}.

For $n=n_e$ even, using the dimensions of the twist operators calculated above, we find
the universal scaling relation
\be
\Tr (\rho_A^{T_{2}})^{n_e}= \frac{d_{n_e}} 
{(\ell_1\ell_2)^{{c}/6({n_e}/2-2/{n_e})} (\ell_1+\ell_2)^{{c}/6({n_e}/2+1/{n_e})}}\,,
\label{3pteven}
\ee
that in the limit $n_e\to1$ gives
\be
|| \rho_A^{T_{2}}||\propto \left(\frac{\ell_1 \ell_2}{\ell_1+\ell_2}\right)^{{c}/4}\qquad \Rightarrow\quad
{\cal E}= \frac{c}4 \ln \frac{\ell_1\ell_2}{\ell_1+\ell_2}+ {\rm cnst}.
\label{3ptevenbis}
\ee
For $n=n_o$ odd we have
\be
\Tr (\rho_A^{T_{2}})^{n_o} =\frac{d_{n_o}}{ (\ell_1\ell_2(\ell_1+\ell_2))^{{c}/{12}(n_o-{1}/{n_o})}} ,
\label{3ptodd}
\ee
that for $n_o\to 1$ gives $\Tr \rho_A^{T_{2}}=1$ as it should.

\subsection{Finite systems.}

All previous results are generalized to the case of a finite system of length $L$ with periodic boundary conditions
by using a conformal mapping from the cylinder (with axis perpendicular to the spatial coordinate) to the plane.
The net effect of the mapping is to replace each length 
 $\ell_i$ with the chord length $(L/\pi) \sin(\pi \ell_i/L)$ in all above formulas.
 
Thus for the case of a pure state in a finite system the generalization of Eqs. (\ref{1inte}) and (\ref{1into}) are
\bea
&&\Tr (\rho_A^{T_2})^{n_e}=
c_{n_e/2}^2 \Big(\frac{L}{\pi a}\sin\frac{\pi \ell}L\Big)^{-{c}/{3}({n_e}/2-2/{n_e})}\,, \label{cftTr1int2}
\\
&&\Tr (\rho_A^{T_2})^{n_o}=
c_{n_o} \Big(\frac{L}{\pi a}\sin\frac{\pi \ell}L\Big)^{-{c}/6(n_o-1/{n_o})}\,, \label{cftTr1int}
\\
&&{\cal E}=\frac{c}2 \ln \Big(\frac{L}{\pi a}\sin\frac{\pi \ell}L\Big)+2\ln c_{1/2}\,.
\label{EN1int}
\eea
For two adjacent intervals embedded in a finite system, the generalizations of Eqs. (\ref{3pteven})  and (\ref{3ptodd}) 
lead to the universal scaling relations
\bea\fl
\Tr (\rho_A^{T_{2}})^{n_e}=\frac{d_{n_e}}{
\left[\frac{L^2}{\pi^2} \sin \big(\frac{\pi \ell_1}L \big) \sin \big(\frac{\pi \ell_2}L\big)\right]^{{c}/6({n_e}/2-2/{n_e})} 
\left[\frac{L}{\pi} \sin \frac{\pi (\ell_1+\ell_2)}L\right]^{{c}/6({n_e}/2+1/{n_e})}}\,,
\nonumber\\ \fl
\Tr (\rho_A^{T_{2}})^{n_o} =\frac{d_{n_o}}{ \left[
\frac{L^3}{\pi^3} \sin \Big(\frac{\pi \ell_1}L\Big)  \sin \Big(\frac{\pi \ell_2}L\Big) \sin \frac{\pi (\ell_1+\ell_2)}L\right]^{{c}/{12}(n_o-1/{n_o})}} ,
\label{3ptfinL}
\eea
and for the logarithmic negativity
\be
{\cal E}=\frac{c}4 \ln\left[ \frac{L}{\pi} \frac{\sin \big(\frac{\pi \ell_1}L\big) \sin \big(\frac{\pi \ell_2}L\big)}{\sin\frac{\pi(\ell_1+\ell_2)}L}\right]
+ {\rm cnst}.
\label{neg3finL}
\ee


\section{Negativity for two disjoint intervals in the Ising universality class}
\label{Sec3}

For a generic CFT with central charge $c$, by using global conformal invariance, 
the four point function of twist fields can be written as \cite{cft-book,fps-08}
\be\fl
\label{4ptgen}
\langle \mathcal{T}_n(z_1)\overline{\cal T}_n(z_2)  
\mathcal{T}_n(z_3)\overline{\cal T}_n(z_4) \rangle
= c_n^2 \left|
\frac{z_{31}  z_{42}}{z_{21} z_{43} z_{41} z_{32}}
\right|^{c/6(n-1/n)}
\mathcal{F}_n(x,\bar{x})\,,
\end{equation}
where the positions $z_i \in \mathbb{C}$, $z_{ij} = z_i-z_j$ and  $x$ is the four-point ratio
\begin{equation}
x \equiv \frac{z_{21} z_{43}}{z_{31} z_{42}}\,,
\ee
and $\bar x$ its complex conjugate. (All distances are measured in units of the UV cutoff $a$.)
The function $\mathcal{F}_n(x,\bar{x})$ is real and must be computed case by case because it depends 
on the full operator content of the theory as explicitly done in some CFTs \cite{fps-08,cct-09,cct-11,ch-09,c-10,h-10,rg-12,hol} 
and also numerically in some lattice models \cite{ffip-08,ip-09,atc-10,fc-10,atc-11,s-12,f-12}, 
but mainly with $x$ real and $0<x<1$.
Since $z_{31} z_{42}/(z_{21} z_{43} z_{41} z_{32}) = 1/(z_{21} z_{43} (1-x))$, 
the normalization of $\mathcal{F}_n(x,\bar{x})$ is given by ${\cal F}_{n}(0,0)=1$. 
Thus, in the limit $x \to 0$ (i.e.  large distance between the two intervals), the four-point function becomes 
the product of two two-point functions normalized through the constant $c_n$.

As explained in Refs. \cite{us-letter,us-long}, in order to give a physical interpretation to Eq. (\ref{4ptgen}) as 
$\Tr \rho_A^n$ or $\Tr (\rho_A^{T_2})^n$, the points $z_i$ must be on the real axis and their order allows to 
distinguish between $\Tr \rho_A^n$ and $\Tr (\rho_A^{T_2})^n$. 
We denote the boundaries of the two intervals with  four real variables $u_j$ and $v_j$ ($j=1,2$)
such that $u_1<v_1<u_2<v_2$ in both cases.
With these definitions,  Eq. (\ref{4ptgen}) for $\Tr \rho_A^n$ corresponds to $z_1=u_1$, $z_2=v_1$, $z_3=u_2$ and $z_4=v_2$,
leading to
\be\fl
\Tr \rho_A^n
=c_n^2 \left(\frac{(u_2-u_1)(v_2-v_1)}{(v_1-u_1)(v_2-u_2)(v_2-u_1)(u_2-v_1)} 
\right)^{{c}/6(n-1/n)} {\cal F}_{n}(x)\,,
\label{Fnd1}
\ee 
where
\begin{equation}
x=\frac{z_{21} z_{43}}{z_{31} z_{42}}=\frac{(v_1-u_1)(v_2-u_2)}{(u_2-u_1)(v_2-v_1)}\,,
\label{4pR}
\end{equation}
and $0<x<1$.  Being $x$ real, the dependence on $\bar{x}=x$ has been dropped. 

Instead, $\Tr (\rho_A^{T_{2}})^{n}$ is obtained from the four-point function (\ref{4ptgen}) with the choice 
$z_1=u_1$, $z_2=v_1$, $z_3=v_2$ and $z_4=u_2$, namely by exchanging $z_3\leftrightarrow z_4$ with respect to the previous case.
The result can be then written as 
\be\fl
\Tr (\rho_A^{T_2})^n
=c_n^2 \left(\frac{(u_2-u_1)(v_2-v_1)}{(v_1-u_1)(v_2-u_2)(v_2-u_1)(u_2-v_1)} 
\right)^{{c}/6(n-1/n)} {\cal G}_{n}(y)\,,
\label{Gn}
\ee 
with
\begin{equation}
y\equiv\frac{z_{21} z_{34}}{z_{41} z_{32}}=\frac{(v_1-u_1)(v_2-u_2)}{(u_2-u_1)(v_2-v_1)}\,,
\label{4pR2}
\end{equation}
and $0<y<1$ and again we dropped the dependence on $\bar y$ because $y$ is real.
Notice that we have chosen the definitions of $x$ and $y$ in such a way that the functional dependence on 
$u_j$ and $v_j$ is the same, but they are two different four-point ratios, as their  dependence on
$z_i$ explicitly shows. The relation between the two is $y=x/(x-1)$.
The expressions (\ref{Fnd1}) and (\ref{Gn}) are related as
\be
{\cal G}_{n}(y)= (1-y)^{{c}/3\left(n-{1}/n\right)}\,{\cal F}_{n}\Big(\frac{y}{y-1}\Big)\,,
\label{GvsF}
\ee
as easily derived from  Eq. (\ref{4ptgen}).

The logarithmic negativity is obtained by considering the even sequence 
$n=n_e$ and taking its analytic continuation  $n_e \to 1$. 
Thus, the functional dependence on $n$ of 
${\cal F}_{n}(x)$ for $x=y/(y-1)<0$ must be different for even and odd $n$, 
in such a way that the limit 
\be
{\cal E}(y)=\lim_{n_e\to1} \ln {\cal G}_{n_e}(y)=
\lim_{n_e\to1} \ln \left[ {\cal F}_{n_e}\Big(\frac{y}{y-1}\Big)\right]\,,
\label{NEG}
\ee
is not trivially equal to $\displaystyle\lim_{n_o\to1} \ln {\cal G}_{n_o}(y)=0$. 
It is natural and convenient to introduce the ratio
\begin{eqnarray}
\label{Rn cft def}
R_n(y) \equiv \frac{\textrm{Tr}(\rho_A^{T_2})^n}{{\rm Tr}\rho_A^n}
&=& \frac{\langle \mathcal{T}_n(u_1)\overline{\cal T}_n(v_1)  
\overline{\mathcal{T}}_n(u_2) \mathcal{T}_n(v_2) \rangle}{\langle \mathcal{T}_n(u_1)\overline{\cal T}_n(v_1)  
\mathcal{T}_n(u_2)\overline{\cal T}_n(v_2) \rangle} 
\\
\rule{0pt}{.65cm}
&=&\frac{{\cal G}_{n}(y)}{{\cal F}_{n}(y)}
=(1-y)^{c/3(n-1/n)} \,
\frac{\mathcal{F}_n(y/(y-1))}{\mathcal{F}_n(y)}\,.
\nonumber
\end{eqnarray}
Indeed, in such ratio the prefactors cancel and a function only of $y \in (0,1)$ is left. Moreover, the ratio in Eq. (\ref{Rn cft def}) 
is an easy quantity to consider in the numerical computations.
Notice that, being $\displaystyle\lim_{n \to 1} \Tr\rho_A^n =1$, the logarithmic negativity is 
obtained equivalently from the replica limit of this ratio
\be
{\cal E}(y)=\ln \lim_{n_e\to1}  R_{n_e}(y)\,.
\ee
Thus, while $\Tr \rho_A^n$ requires ${\cal F}_{n}(x)$ only for $x \in (0,1)$, in order  to extract $\Tr (\rho_A^{T_2})^n$, and consequently the negativity ${\cal E}(y)$, one has to compute the function ${\cal F}_{n}(x)$ also for $x <0$.

We remark that for the special case $n=2$, we have $\overline{\mathcal{T}}_2= \mathcal{T}_2$
because $\mathcal{T}_2^2$ is the identity operator.  This implies that $R_2(y) =1$ for $0<y<1$  identically.

In order to find $\mathcal{F}_n(x,\bar{x})$ for the Ising model, it is useful to review the corresponding result for the free compactified boson (i.e. a Luttinger liquid field theory), derived  in Ref. \cite{us-long}.


\subsection{Free compactified boson}

For the free {\it real} boson compactified on a circle of radius $R$, the function $\mathcal{F}_n(x,\bar{x})$ 
in Eq. (\ref{4ptgen}) for any complex $x$ has been computed in \cite{us-long}. 
This has been possible starting from some results in Refs.  \cite{dixon,z-87} and generalizing to 
$x \in \mathbb{C}$ the procedure employed in Ref. \cite{cct-09} for $0<x<1$.
The final result is \cite{us-long}
\begin{equation}
\label{Fn boson}
\mathcal{F}_n(x,\bar{x})
=
\frac{\Theta\big({\bf 0} | T_\eta(x,\bar{x}) \big)}{\prod_{k=1}^{n-1} |F_{k/n}(x)|}\,,
\end{equation}
where we introduced $ F_{k/n}(x) \equiv\, _2F_1(k/n,1-k/n;1;x)$ to denote the special case of the hypergeometric function 
$_2F_1(a,b;c;z)$.
Here $\Theta({\bf 0}|K)$ is of the Riemann-Siegel theta function which is generically defined as 
\begin{equation}
\label{theta Riemann def}
\Theta({\bf 0}|K)\,\equiv\,
\sum_{{\bf m} \, \in\,\mathbf{Z}^{p}}
\exp\big[i\pi\,{\bf m}\cdot K \cdot {\bf m}\big]\,,
\end{equation}
where $K$ is a $p \times p$ symmetric complex matrix with positive definite imaginary part and 
${\bf 0} $ is the $p$ dimensional vector made by zeros. 
The $2(n-1) \times 2(n-1)$ symmetric complex matrix $T_\eta(x,\bar{x})$ entering in Eq. (\ref{Fn boson}) is 
\be
 T_\eta(x,\bar{x}) =
 \left(\begin{array}{cc}
 i \eta \mathcal{I} & \mathcal{R} \\ 
  \mathcal{R} &  i \mathcal{I} /\eta
 \end{array}
 \right)\,,
\ee
where the parameter $\eta$ is proportional to $R^2$ and the  $(n-1) \times (n-1)$ symmetric real matrices $\mathcal{R}$ 
and $ \mathcal{I}$ are respectively the real and the imaginary part of the $(n-1) \times (n-1)$ period matrix 
\be
\label{tau def}
\tau(x) = \mathcal{R}+ i \,\mathcal{I}
= \frac{2}{n} \sum_{k=1}^{n-1} \tau_{k/n}(x) \sin(\pi k/n)\, C_{k/n}\,,
\ee
with
\be
\label{taukn def}
\tau_{k/n}(x) \equiv  i\, \frac{F_{k/n}(1-x)}{F_{k/n}(x)}\,,
\ee
and $C_{k/n}$ is the $(n-1) \times (n-1)$ symmetric matrix whose elements read 
\be
\big(C_{k/n}\big)_{rs} = \cos\big[2\pi k/n(r-s)\big]\,,
\hspace{1cm} r,s= 1, \dots , n-1\,.
\ee
Notice that, because of the $k \leftrightarrow n-k$ invariance of the functions $F_{k/n}$, for $n$ odd the denominator in (\ref{Fn boson}) becomes the square of a product over $k$ going from 1 to $(n_o-1)/2$.
Similarly, (\ref{tau def}) can be written as a sum over $k$ going from 1 to $(n_o-1)/2$. 
This suggests that the term corresponding to $k/n=1/2$, occurring only for even $n$, plays a key role in the limit (\ref{NEG}), 
as shown explicitly in \cite{us-long} for the non-compactified boson in the limit of close intervals.

Few comments are now in order regarding Eq.~(\ref{Fn boson}) before proceeding to the partial transpose.
Comparing Eq. (\ref{Fn boson}) with old results about CFT on higher genus Riemann 
surfaces \cite{z-87, dvv-87, AlvarezGaume-86, Knizhnik-87, AlvarezGaume-87}, one observes 
that $\mathcal{F}_n(x,\bar{x})$ is related to the partition function 
$\mathcal{Z} = \mathcal{Z}^{ \rm qu}  \mathcal{Z}^{ \rm cl}(R)$ 
on a genus $g=n-1$ Riemann surface, which can be factorized into a quantum part and a classical part, 
and all the dependence on $R$ of $\mathcal{Z}$ is 
contained in the latter. In particular, one finds that
$ \mathcal{Z}^{\rm cl }(R)  = \Theta ({\bf 0} | T_\eta(x,\bar{x}))$ and the period matrix of the Riemann surface is 
$\tau(x)$ given in (\ref{tau def}). As already remarked \cite{dixon, cct-09}, here we are not dealing with generic genus $g$ 
Riemann surfaces, but with a subclass of them obtained through the replica method (see the appendix of \cite{cct-11} 
for a pictorial representation).
Thus, while for $\Tr \rho_A^n$ one needs the period matrix $\tau(x)$ only for $0<x<1$, where its real part $\mathcal{R}$ 
vanishes identically, for $\Tr (\rho_A^{T_2})^n$ the period matrix $\tau(x)$ for $x<0$ is required and in this regime 
$\mathcal{R}$ enters in a crucial way.
The expression (\ref{Fn boson}) is explicitly invariant under $\eta \leftrightarrow 1/\eta$. We recall that one can also rewrite 
Eq. (\ref{Fn boson}) in a form which makes manifest its invariance under  $x \leftrightarrow 1-x$. This form is 
particularly useful also to study the regime $\eta \to \infty$ of non-compactified boson because 
some simplifications occur and the Riemann-Siegel theta function does not contribute in this limit \cite{us-long}.

At this point, it is easy to write $\Tr (\rho_A^{T_2})^n$ for the compactified boson from Eq. (\ref{Gn}), 
where ${\cal G}_n(y)$ is given by (\ref{GvsF}) with $c=1$ and ${\cal F}_n(x)$ by (\ref{Fn boson}), i.e.
\be
\label{Gn boson}
{\cal G}_n(y)  =(1-y)^{(n-1/n)/3} \,
\frac{\Theta\big({\bf 0} | T_\eta(\frac{y}{y-1})\big)}{ \prod_{k=1}^{n-1}  \big| F_{k/n}(\frac{y}{y-1}) \big| }\,,
\ee
where $0<y<1$.  From Eqs. (\ref{Gn boson}) and (\ref{Fn boson}), we can write the ratio (\ref{Rn cft def}) for the free compactified boson
\begin{equation}
\label{Rn boson def}
R_n(y)  =(1-y)^{(n-1/n)/3} 
 \left(\,
\prod_{k=1}^{n-1} \bigg| \frac{F_{k/n}(y)}{F_{k/n}(\frac{y}{y-1})} \bigg|
\,\right)
\frac{\Theta\big({\bf 0} | T_\eta(\frac{y}{y-1})\big)}{\Theta\big({\bf 0} | T_\eta(y) \big)}\,.
\end{equation}
In the decompactification regime $\eta\to\infty$, this formula has been checked numerically 
in Ref. \cite{us-long} for an harmonic chain with periodic boundary conditions, whose Hamiltonian is the lattice discretization 
of a free boson (Klein-Gordon action). 
Always in this regime and in the limit $y \rightarrow 1^-$ of close intervals, the analytic continuation $n_e \to 1$ 
of the even sequence $R_{n_e}(y)$ has been explicitly carried out  \cite{us-long}. 
The analytic continuation $n_e \rightarrow 1$ for finite and generic $\eta$ is still unknown. A similar problem occurs for the 
von Neumann entanglement entropy through the analytic continuation of the R\'enyi entropies of two disjoint intervals.
We finally mention that the results for some finite $\eta$ have been recently checked in Monte Carlo simulations
\cite{vinc-new}.


\subsection{Ising conformal field theory}

In the previous section we have shown that ${\cal F}_n(x,\bar{x})$ can be found from the partition function of the 
corresponding model on the genus $g=n-1$ Riemann surface obtained from the replica method.
The partition function of the Ising model on a genus $g$ Riemann surface has been derived in the eighties  
\cite{dvv-87, AlvarezGaume-86, AlvarezGaume-87, Bernard-88} by considering $\mathbb{Z}_2$ orbifolds 
models on generic Riemann surfaces. 
From the knowledge of the partition function 
$\mathcal{Z} = \mathcal{Z}^{ \rm qu}  \mathcal{Z}^{ \rm cl}(R) $ of the compactified boson on the same 
Riemann surface, we can explicitly access the period matrix on the Riemann surface. 
Then, using the mentioned old results for
the Ising model on a genus $g$ Riemann surface, we can finally write \cite{cct-11}
\be
\mathcal{Z}_{ \rm Ising } = (\mathcal{Z}^{ \rm qu})^{1/2} 
\,2^{-g}
\sum_{{\bf e}} 
\big|
\Theta[{\bf e}]\big({\bf 0}| \tau\big)
\big|\,,
\ee
where $\tau$ is the $g \times g$ period matrix. Here we introduced the Riemann-Siegel theta function 
with characteristic ${\bf e}$, which is defined as 
\be
\label{RS theta def}
\Theta[{\bf e}]({\bf z}| \tau) 
=
\sum_{{\bf m}\, \in\, \mathbb{Z}^{g}}
e^{i\pi \big[({\bf m}+  \boldsymbol{\varepsilon}) \cdot \tau \cdot ({\bf m}+  \boldsymbol{\varepsilon})
+2({\bf m}+  \boldsymbol{\varepsilon})\cdot  ({\bf z} + \boldsymbol{\delta})\big]}\,,
\ee
where ${\bf z} \in \mathbb{C}^g/(\mathbb{Z}^g + \tau \,\mathbb{Z}^g)$ is a $g$ dimensional complex vector and the characteristic ${\bf e}$ is given by a pair of $g$ dimensional vectors $ \boldsymbol{\varepsilon}$ and $ \boldsymbol{\delta}$ made by $0$'s and $1/2$'s
\be
[ {\bf e}  ]
= \bigg[\begin{array}{c}
 \boldsymbol{\varepsilon}\\
 \boldsymbol{\delta}
 \end{array}\bigg]
= \bigg[\begin{array}{c}
\varepsilon_1, \dots, \varepsilon_{g} \\
\delta_1, \dots, \delta_{g} 
\end{array}\bigg]\,,
\hspace{1.6cm}
\varepsilon_i, \delta_i \in \{0,1/2\}\,.
\ee
In the case at hand, the period matrix $\tau(x)$ is given by Eq. (\ref{tau def}) and the partition function for the compactified boson 
$\mathcal{Z} = \mathcal{Z}^{\rm qu}  \mathcal{Z}^{ \rm cl}(R) $  is Eq. (\ref{Fn boson}). 
Thus, for the Ising model we have
\be
\label{Fn ising}
\mathcal{F}_n(x,\bar{x})
=
\frac{\sum_{{\bf e}} 
\big|
\Theta[{\bf e}]\big({\bf 0}| \tau(x)\big)
\big|}{2^{n-1}\prod_{k=1}^{n-1} \big| F_{k/n}(x)\big|^{1/2}}\,.
\ee
For $0<x<1$, this function reduces to $\mathcal{F}_n(x)$ found in \cite{cct-11} for the R\'enyi entropies of the Ising model 
and it has been tested numerically through various methods \cite{atc-10, fc-10, atc-11}, finding agreement once 
the finite size corrections are properly taken into account.

\begin{figure}[t]
\begin{center}
\includegraphics[width=.79\textwidth]{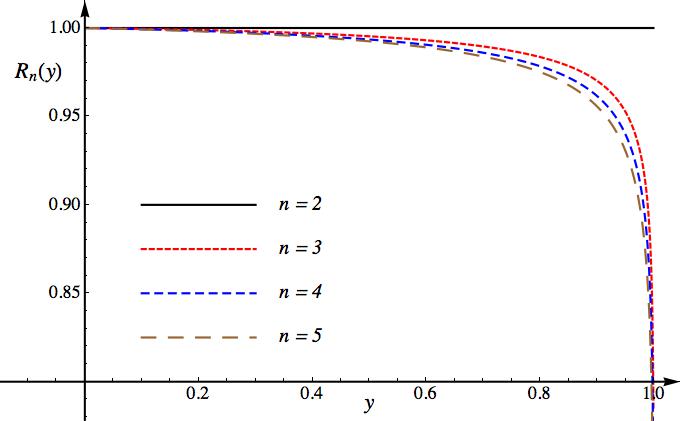}
\end{center}
\caption{The function $R_n(y)$ for the Ising conformal field theory as a function of $y$
for several values of $n$ ($n=2,3,4,5$ curves from top to bottom). 
}
\label{RnCFT}
\end{figure}

Writing $\Tr (\rho_A^{T_2})^n$ for the Ising model is now easy by using Eq. (\ref{Gn}) with $c=1/2$, 
and ${\cal G}_n(y) $ from Eqs. (\ref{GvsF})  and (\ref{Fn ising}), namely
\be
\label{Gn ising}
{\cal G}_n(y) 
=(1-y)^{(n-1/n)/6}
\,\frac{\sum_{{\bf e}} 
\big| \Theta[{\bf e}]\big({\bf 0}| \tau(\frac{y}{y-1})\big) \big|}{ 2^{n-1} \prod_{k=1}^{n-1} \big| F_{k/n}(\frac{y}{y-1}) \big|^{1/2}} \,,
\ee
where $0<y<1$ and $\tau$ is given by Eq. (\ref{tau def}). 
From this expression and Eq. (\ref{Fn ising}), we can write the ratio (\ref{Rn cft def}) for the Ising model\footnote{We 
have been informed that this result has been derived independently also by V. Alba \cite{vinc-new}.}
\be\fl
\label{Rn ising def}
R_n(y) 
=(1-y)^{(n-1/n)/6}\,
\prod_{k=1}^{n-1} \bigg| \frac{F_{k/n}(y)}{F_{k/n}(\frac{y}{y-1})} \bigg|^{{1}/{2}}\,
\frac{\sum_{{\bf e}} 
\big| \Theta[{\bf e}]\big({\bf 0}| \tau(\frac{y}{y-1})\big) \big|}{
\sum_{{\bf e}} 
\big| \Theta[{\bf e}]\big({\bf 0}| \tau(y)\big) \big|}\,.
\ee
The curves $R_n(y)$ for $0<y<1$ are shown in Fig. \ref{RnCFT} for $n=2,3,4,5$. 
They are concave curves with $R_n(0)=1$, 
which stay below 1 for $n>2$. 
When $n=2$, we have $R_{2}(y) =1$ identically, as we will explicitly prove below. 
Moreover, $R_{n_2}(y) < R_{n_1}(y)$ when $n_2 > n_1$, for all $0<y<1$.
Approaching $y=1$, $R_n(y)$ approaches zero very steeply in a calculable power-law way 
that, for a general model, depends only on the central charge \cite{us-long}.

It is worth mentioning that after a superficial analysis, 
one could erroneously conclude that the sum over the characteristics occurring in Eq.
(\ref{Fn ising}) involves $2^{2g}$ terms. 
This is not true because many of these terms are identically zero.
In order to see this, let us introduce the parity of the characteristic ${\bf e}$ which is the parity of the integer
number $4  \boldsymbol{\varepsilon}\cdot \boldsymbol{\delta}$ which is related to 
the transformation property of the Riemann-Siegel theta function  \cite{AlvarezGaume-86}
\be
\label{theta parity}
\Theta[{\bf e}](- {\bf z}| \tau)  = 
(-1)^{4  \boldsymbol{\varepsilon}\cdot \boldsymbol{\delta}}
\Theta[{\bf e}]({\bf z}| \tau)\,.
\ee
There are $2^{g-1}(2^g+1)$ even characteristics and $2^{g-1}(2^g -1)$ odd ones. 
Eq. (\ref{theta parity}) implies that 
$\Theta[{\bf e}]({\bf 0}| \tau) =0$ identically for odd characteristics.
This means that the sum over the characteristics which defines $\mathcal{F}_n(x,\bar{x})$ in (\ref{Fn ising}) is made only 
by $2^{n-2}(2^{n-1}+1)$ terms, namely the ones with even ${\bf e}$.

\subsection{The special case $n=2$ for Ising model.}

For the special case of $n=2$,  since $\mathcal{T}_2=\overline{\cal T}_2$, we have $R_2(y) =1$ identically in $0<y<1$. 
This is true for any CFT, as already noted above, and therefore also for the Ising model. 
Nevertheless, we find it instructive to see explicitly how this simplification occurs in  Eq. (\ref{Rn ising def}), 
since this is not apparent from the formula and the  explicit proof is related to the modular invariance 
of the torus partition function \cite{mod-inv}.

When $n=2$ the Riemann surface is a torus and the period matrix reduces to its modulus $\tau(x) =\tau_{1/2}(x)$, 
which is a complex number. In this case 
the four-point ratio $x \in \mathbb{C}$ and its modulus $\tau(x)$ are related as  \cite{dixon,fps-08}
\be\fl
\label{n=2 basics}
F_{1/2}(x) = \frac{2}{\pi} K(x) = \theta_3(\tau(x))^2,
\qquad
\tau(x) = \frac{i K(1-x)}{K(x)},
\qquad
x(\tau) = \left( \frac{\theta_2(\tau)}{\theta_3(\tau)} \right)^4,
\ee
where $K(x)$ is the complete elliptic integral of the first kind. From the properties of $K$, for $0<y<1$ one finds \cite{Bateman}
\begin{eqnarray}
\label{2F1 n=2 transform up}
F_{1/2}\left(\frac{1}{1-y}\right) & =&  \frac{2}{\pi} \sqrt{1-y} 
\big( K(1-y) - i K(y) \big)\,,
\\
\rule{0pt}{.65cm}
\label{2F1 n=2 transform down}
F_{1/2}\left(\frac{y}{y-1}\right) & =&\frac{2}{\pi} \sqrt{1-y}\, K(y)\,. 
\end{eqnarray}
Given $\tau = \tau(y) $ in the second formula of (\ref{n=2 basics}), (\ref{2F1 n=2 transform up}) and (\ref{2F1 n=2 transform down}), it is straightforward to observe that for $0<y<1$ we have
\begin{equation}
\label{tauprime def}
\tau' \equiv \tau\left(\frac{y}{y-1}\right) = \tau+1 \,.
\end{equation}
We also need to recall the following transformation properties of the Jacobi theta functions under the modular transformation $\tau \rightarrow \tau +1$  \cite{cft-book,mod-inv}
\be\fl
\label{jacobi tau+1}
\theta_2(\tau+1) = e^{i\pi/4} \theta_2(\tau)\,,
\hspace{1.2cm}
\theta_3(\tau+1) = \theta_4(\tau)\,,
\hspace{1.2cm}
\theta_4(\tau+1) = \theta_3(\tau)\,.
\ee
From the first formula in (\ref{n=2 basics}), (\ref{tauprime def}) and the second formula in (\ref{jacobi tau+1}), one finds that
\begin{equation}
\label{Ftheta4 identity}
F_{1/2}\left(\frac{y}{y-1}\right)  = \theta_3(\tau+1)^2 =  \theta_4(\tau)^2\,.
\end{equation}
Given the expressions above, we are ready to prove that $R_2(y) =1$ for $0<y<1$. 
Indeed, from Eqs. (\ref{n=2 basics}) and (\ref{Ftheta4 identity}) we get
\be
(1-y)^{1/4} = \left| \frac{\theta_4(\tau)}{\theta_3(\tau)} \right|\,,
\qquad
\bigg| \frac{F_{1/2}(y)}{F_{1/2}(\frac{y}{y-1})} \bigg|^{1/2}
= \left| \frac{\theta_3(\tau)}{\theta_4(\tau)} \right|\,,
\ee
where in the first equation the identity $\theta^4_4 = \theta^4_3- \theta^4_2$ has been employed.
As for the ratio containing the Riemann-Siegel theta functions in (\ref{Rn ising def}), when $n=2$ and $0<y<1$, it simplifies to
\begin{equation}
\frac{|\theta_2(\tau')|+|\theta_3(\tau')|+|\theta_4(\tau')|}{|\theta_2(\tau)|+|\theta_3(\tau)|+|\theta_4(\tau)|} 
= 1\,,
\end{equation}
where Eq. (\ref{tauprime def}) and the modular transformations (\ref{jacobi tau+1}) have been employed.


\section{The partial transposition with Tree Tensor Networks}

\label{Sec4}

The Ising chain in a transverse magnetic field is the most studied one dimensional model 
and this is mainly due to fact that  it is a non-trivial model that can be 
mapped to a system of free fermions by means of a Jordan-Wigner transformation \cite{sach-book}. 
However, it is still a very hard problem to find an effective way to calculate the 
partial transpose of the reduced density matrix (or at least its eigenvalues) in the free fermions formulation as opposite 
to models of free bosons 
for which many results are already available (see e.g. \cite{Audenaert02,Neg2,us-long}).
For this reason, in order to check the analytical predictions from conformal field theory,
we resort to purely numerical methods based on tree tensor techniques which 
already have been very effective in calculating of the entanglement entropies of two disjoint intervals 
\cite{atc-10,atc-11}.
An alternative approach based on classical Monte Carlo 
simulations (generalizing the method in Refs. \cite{cd-08,atc-10,atc-11,gt-11}) has been also  independently
developed at the same time of this work by V. Alba \cite{vinc-new}. 

\begin{figure}
\rule{0pt}{.5cm}
\begin{center}
 \includegraphics[width=0.9\textwidth]{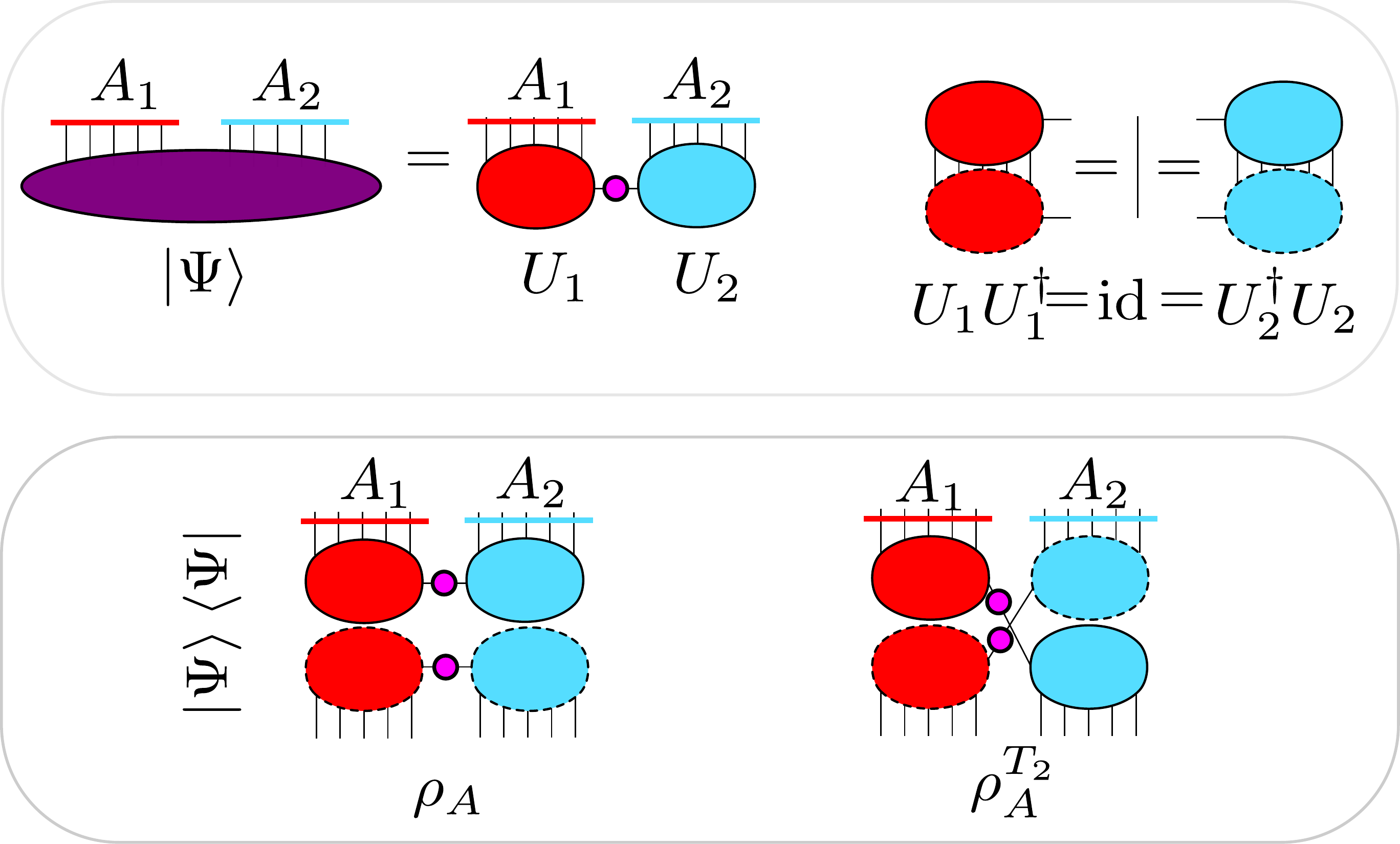}
 \end{center}
\caption{Graphical representations of some quantities for a pure state and of some useful properties 
in  the language of the tensor networks. 
Top: We consider the bipartition of the pure state $|\Psi\rangle$  in $A_1$ and $A_2$ and its Schmidt decomposition 
through the isometries $U_1$ and $U_2$. 
Bottom left: the density matrix $\rho_A$. 
Bottom right: the partial transpose $\rho_A^{T_2}$.
\label{fig:def}}  
\end{figure}

\subsection{Tensor networks: notation and examples}

We consider a 1D lattice $\mathcal{L}$ made of $L $ sites and on each site $s\in \mathcal{L}$ 
we set a local Hilbert space $\mathbb{V}_s$ of finite dimension $d$ (e.g. for 
a spin-$1/2$ for which $d=2$). 
An arbitrary pure state $\ket{\Psi} \in \mathbb{V}^{\otimes L}$ defined on the lattice $\mathcal{L}$, 
in the local basis $\{\ket{1_s}, \ket{2_s}, \cdots, \ket{d_s}\}$ of $\mathbb{V}_s$, 
can be expanded as follows
\begin{eqnarray}
	|\Psi\rangle = \!\sum_{i_1=1}^d ~ \sum_{i_2=1}^d \cdots \sum_{i_L=1}^d 
	T_{i_1i_2 \cdots i_L} \ket{i_1} \ket{ i_2} \cdots \ket{i_L},
\label{eq:local_expansion}
\end{eqnarray}
i.e. can be encoded in a tensor $T$ whose
$d^{L}$ elements $T_{i_1i_2 \cdots i_L} \in \mathbb{C}$ fully determine the state. 
We refer to the index $1 \leqslant i_s \leqslant d$, labeling a local basis for site $s$, as a \emph{physical} index. 

The tensor network approach \cite{tns} is a powerful way to rewrite the exponentially large tensor $T$ 
in Eq. (\ref{eq:local_expansion}) as a combination of smaller tensors.
However, the manipulations required to make the products of these tensors 
easily become too long and cumbersome to write explicitly all indices and   
then it is very convenient to represent them by diagrams.
In these diagrams each tensor is 
represented by  circles having some outgoing lines and 
each of them represents a tensor's index.
For each tensor, we denote its complex conjugate with the same circle 
delimitated by a dashed line instead of a continuos one. 
A line shared by two tensors  represents a contraction of
 a particular index.

\begin{figure}
\vspace{1cm}
\begin{center}
 \includegraphics[width=.98\textwidth]{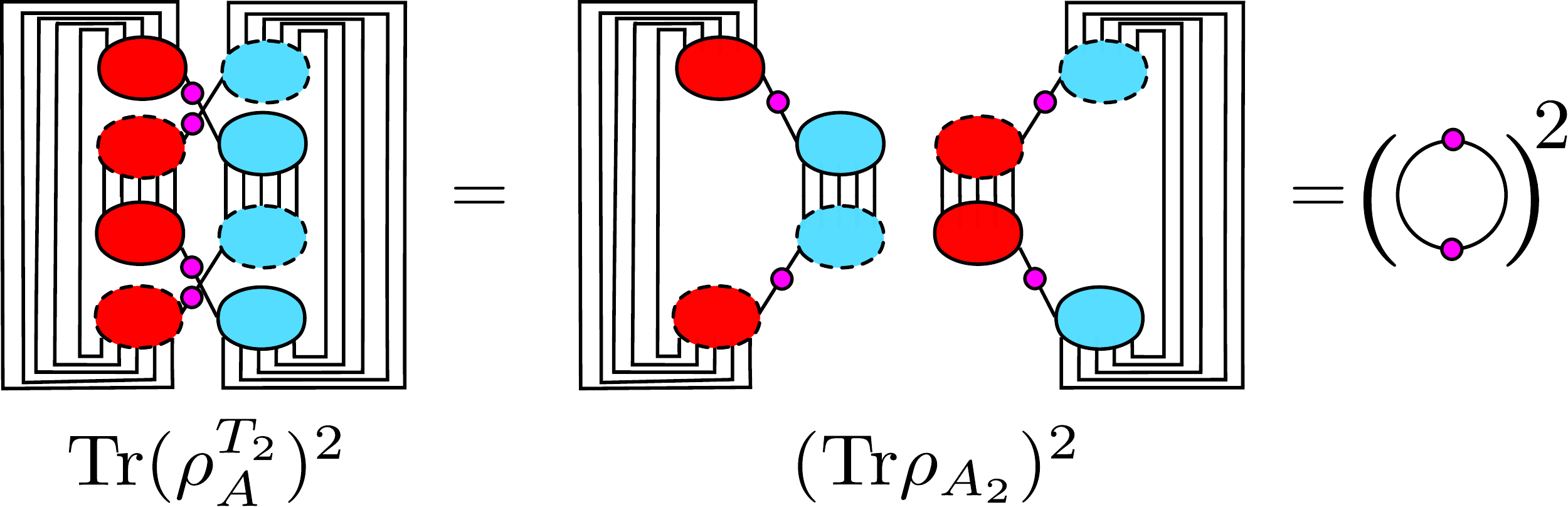}
  \end{center}
\caption{Graphical representation of the identity $\Tr (\rho_A^{T_2})^{n_e}=(\Tr\rho_{A_2}^{n_e/2})^2$ for $n_e=2$.
\label{fig:even}}  
\end{figure}

In order to show the power and simplicity of this diagrammatic representation in a concrete example, 
we will graphically prove the identities $\Tr (\rho_A^{T_2})^{n_e}=(\Tr\rho_{A_2}^{n_e/2})^2$ 
and $\Tr (\rho_A^{T_2})^{n_o} =\Tr \rho_{A_2}^{n_o}$ \cite{us-long}, 
which hold when $B=\emptyset$ and the whole chain $A=A_1 \cup A_2$ is in a pure state.
Let us consider the case of a chain $A$ made by $8$ spins and a bipartition such that both $A_1$ and $A_2$ contain $4$ spins
(but the validity of the proof does not rely on these numbers).
The pure state $|\Psi\rangle$ is depicted in the top of Fig. \ref{fig:def} where also its Schmidt decomposition is shown 
(the different colors of the circles refer to $A_1$ and $A_2$).
The small (purple) circle on the line connecting the two big circles
stands for the diagonal matrix formed by the Schmidt coefficients $c_\alpha$, i.e.  
$|\Psi\rangle=\sum_\alpha c_\alpha |e^{(1)}_\alpha \rangle\otimes  |e^{(2)}_\alpha \rangle$.
We recall that the Schmidt vectors $ |e^{(1)}_\alpha \rangle$ and $ |e^{(2)}_\alpha \rangle$
provide an orthonormal basis and this ensures that the
transformation matrices $U_1$ and $U_2$ are {\it isometric} \cite{shividal06}, a property that 
graphically can be depicted as in Fig. \ref{fig:def} (top right).
The  density matrix $\rho_A=|\Psi \rangle \langle \Psi|$ is given by the picture on the bottom (left part) of Fig. \ref{fig:def}. 
The definition of $\rho_A^{T_2}$ given in Eq. (\ref{rhoAT2def}) can be implemented in the  tensor network language by 
interchanging the tensor associated to $A_2$ and its complex conjugate, as again shown in Fig. \ref{fig:def} (bottom right).

Now we are ready to show how the above identities for pure states can be easily proved 
by using the graphical representation. 
We first consider even powers, i.e. $\Tr (\rho_A^{T_2})^{n_e}=(\Tr\rho_{A_2}^{n_e/2})^2$. 
In Fig. \ref{fig:even}, for sake of graphical simplicity, we restrict to $n_e =2$  (which  trivially gives $1$ 
because of the normalization condition $\Tr\rho_{A_2}=1$) but the key steps hold for any even $n_e$.
$\Tr (\rho_A^{T_2})^{2}$ is  obtained by contracting the indices 
of two copies of $\rho_A^{T_2}$ as shown in Fig. \ref{fig:even} (left), and then realizing that the result 
factorizes into the product of two $\Tr\rho_{A_2}$  as in the middle of Fig. \ref{fig:even}. 
Using the isometry property  in Fig. \ref{fig:def},  each $\Tr\rho_{A_2}$ further reduces to the sum of the squared Schmidt 
coefficients, as it should.
 $\Tr (\rho_A^{T_2})^{n_o}$ is graphically calculated in Fig.  \ref{fig:odd} for $n_o=3$.
As above, we first write $\Tr (\rho_A^{T_2})^3 $ through the proper contractions of three copies of $\rho_A^{T_2}$. 
Then,  by rearranging the order of the tensors, this can be put as in the center of Fig.  \ref{fig:odd},
which is $\Tr \rho_{A_2}^{3}$. 
Again, the last step can be obtained through the isometry property in Fig. \ref{fig:def}.
These two examples make manifest the power of the graphical notation. 
Finally, it is worth stressing, that the same proof could have been done in any other basis.
We choose the Schmidt one only because in this basis the reduced density $\rho_{A_2}$ matrix is diagonal.

\begin{figure}
\vspace{1cm}
\begin{center}
\includegraphics[width=0.8\textwidth]{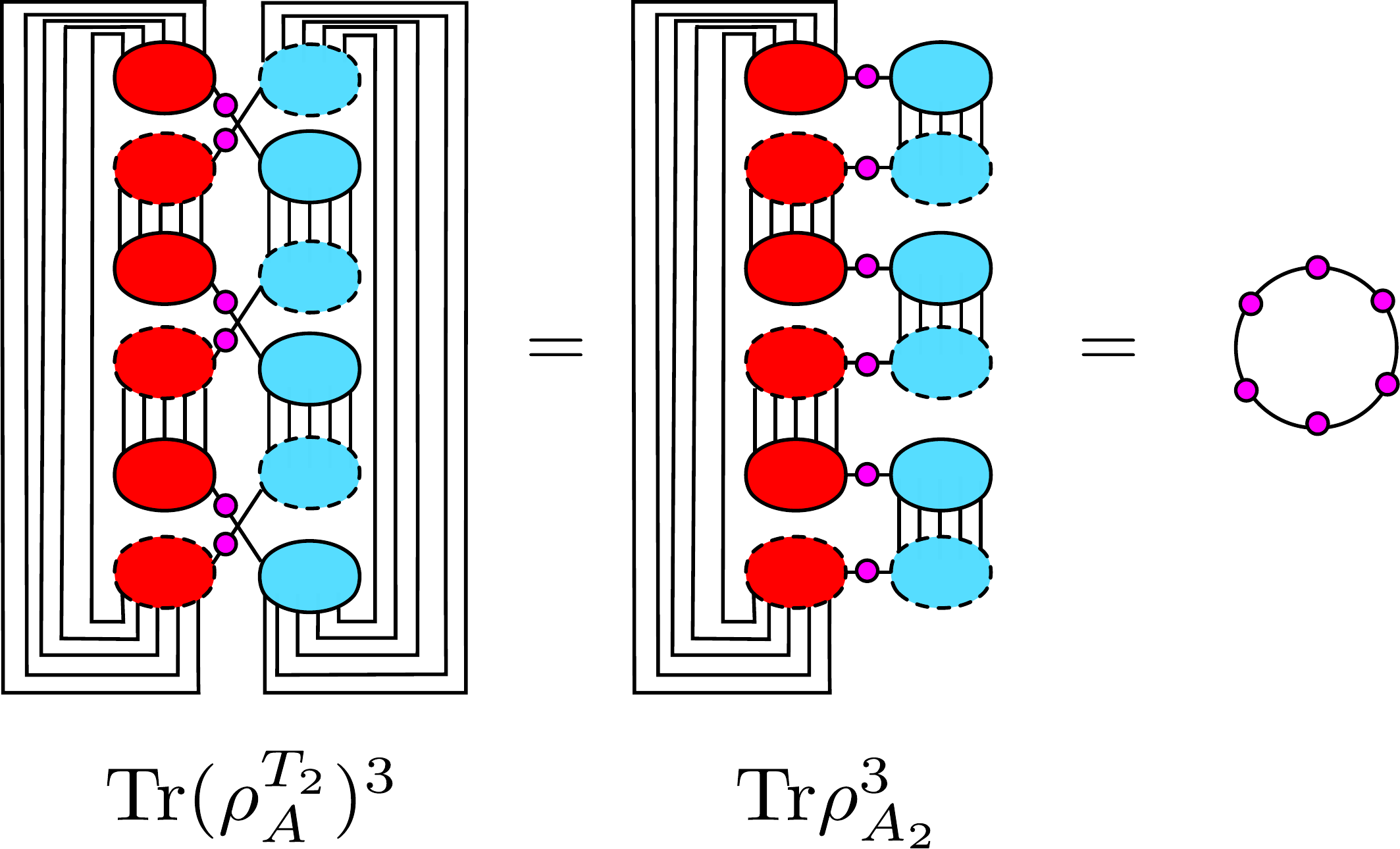}
 \end{center}
\caption{Graphical representation of the identity $\Tr (\rho_A^{T_2})^{n_o} =\Tr \rho_{A_2}^{n_o}$ for $n_o=3$. 
\label{fig:odd}}  
\end{figure}

\subsection{The Tree Tensor Network}

Hereafter we focus on the specific tensor network that we have used for the calculation of the partially transposed reduced density 
matrix, i.e. the Tree Tensor Networks (TTN). 
We only  briefly recall the basics of the method and for a more detailed explanation, we refer the reader to Refs. \cite{gt-11,tns,shividal06,tev-09,fannes92,friedman97,lepetit00,delgado02,nagaj08,silvi09,hubener10,murg10,hubener11,tv-11}.

Representing the pure state $|\Psi\rangle $ as  in Eq. (\ref{eq:local_expansion}) with a TTN implies 
that the $d^L$ coefficients of the tensor  $T_{i_1i_2 \cdots i_L}$ can be obtained as  the result of the contraction 
of a network made by tensors $w$ (built with a smaller number of indices) forming a pattern whose shape 
reminds a tree structure, as graphically 
shown in Fig. \ref{fig:ttn} for a lattice of $L=16$ sites and tensors $w$ which have three indices. 
The TTN decomposition of  $T_{i_1i_2 \cdots i_L}$ is given by a collection of tensors $w$, hierarchically organized  in layers labelled by $\tau$, with $\tau=0,\dots, N$ and $(N+1)=\log_2 L $.  
The tensors $w$ have both \emph{bond} indices and {\it physical} indices. 
All the bond indices are contracted while the $L$ physical indices are left non contracted, 
so that they can be thought as the leaves of the tree.
In this way, the TTN encodes the $d^L$ complex coefficients $T_{i_1i_2 \cdots i_L}$ of the 
state in Eq. (\ref{eq:local_expansion}).

\begin{figure}
\vspace{.4cm}
\begin{center}
\includegraphics[width=0.9\textwidth]{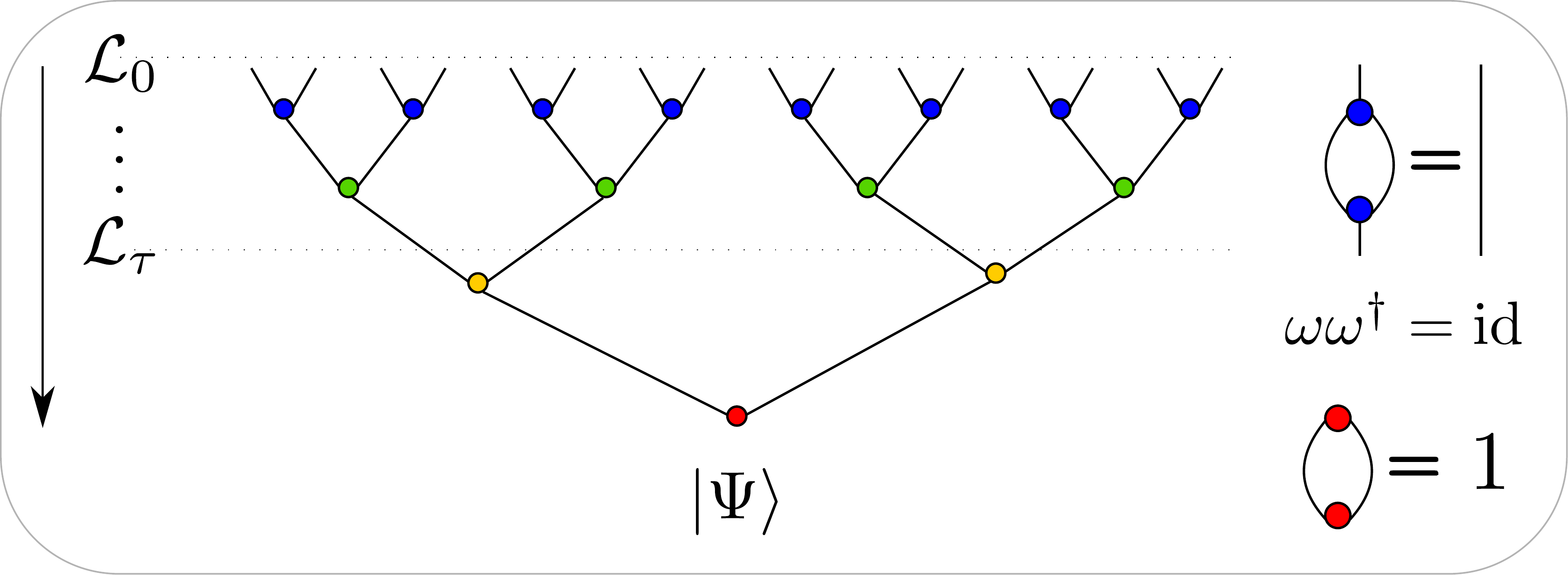}
\end{center}
\caption{ A state $|\Psi \rangle $ encoded by a tree tensor network involving 16 spins.
\label{fig:ttn}}  
\end{figure}

In the example of Fig.  \ref{fig:ttn}, each elementary tensor $w$ has at most one lower index (leg) $\alpha$ and two upper 
indices (legs) $\beta_1$ and  $\beta_2$ and for this reason is called a {\it binary tree}, which is the only 
structure we will consider.  
At each level $\tau$, we allow the indices $\beta_i$ to run from $1$ to $\chi_\tau$.
Without loss of generality, the tensor $w$ can be chosen to be isometric \cite{shividal06}, i.e. 
$\omega \omega^\dagger =\textrm{id}_\chi$, with $\textrm{id}_\chi$ being the $\chi\times \chi$ identity tensor.
Explicitly, this condition reads 
\begin{equation}\fl
\sum_{\beta_1 , \beta_2=1}^{\chi_\tau} (w)^{\beta_1 \beta_2}_{\alpha}(w^{\dagger})_{\beta_1 \beta_2}^{\alpha'} = \delta_{\alpha\alpha'}\,,
\hspace{1cm}
\tau=0, \dots, N-1\,,
\hspace{.5cm}
\alpha,\alpha' = 1,\dots, \chi_{\tau + 1}\,.
\label{eq:isometry}
\end{equation}
The bottommost tensor of the network (the red one in Fig.  \ref{fig:ttn}), i.e. the only tensor at the layer $N$, 
is different from the others. Indeed, since it does not possess any lower index, it represents a normalized vector, namely
\begin{eqnarray}
\sum_{\beta_1 , \beta_2} (w)^{\beta_1 \beta_2}(w^{\dagger})_{\beta_1 \beta_2} = 1\, ,
\hspace{1.5cm}
\tau=N\,.
\label{eq:up-vect}
\end{eqnarray}
The properties (\ref{eq:isometry}) and (\ref{eq:up-vect}), which are graphically shown on the right of Fig.  \ref{fig:ttn}, 
guarantee that the state $|\Psi\rangle$ in (\ref{eq:local_expansion}) is normalized to 1, i.e. $\braket{\Psi | \Psi}=1$.
These considerations can be generalized to cases with elementary tensors having more upper and lower legs \cite{tev-09}.
The natural layered structure of the TTN is emphasized by the arrow on the left of Fig.  \ref{fig:ttn}.
Each layer is an isometric transformation that  maps a lattice  $\mathcal{L}_\tau$ consisting of $L_\tau$ sites 
to a lattice  $\mathcal{L}_{\tau+1}$ with $L_{\tau+1}=L_\tau/2$ sites. 
The physical lattice is $\mathcal{L}_{0}$, whose length $L_0=L$ must be a power of $2$. 
The arrow on the left of Fig.  \ref{fig:ttn} shows the direction along which the coarse-graining of the lattice increases.

In order to describe the ground state of a local Hamiltonian $H$, the tensors of the network should describe  
the state $| \Psi \rangle $ that minimizes $\langle \Psi | H | \Psi \rangle $. 
This can be achieved using the (numerical) algorithm described in Ref. \cite{tev-09} whose
computational cost is 
$\chi_{max}^4$, where $\chi_{max}=\textrm{max}_\tau\{\chi_\tau\}$ is  the largest
bond dimension  in the network. 
Recently, the TTN has been combined with Monte Carlo sampling obtaining an algorithm whose cost scales with 
 $\chi_{max}^3$ \cite{ferrisvidal11}.
Each layer of the  TTN encoding the ground state of $H$ is a coarse-graining transformation that selects those states of $ \mathcal{L}_{\tau}$ which are relevant for the low energy physics of $H$ at the scale of $\mathcal{L}_{\tau+1}$. 
When  both the Hamiltonian and its ground state are  translational invariant,   we can  force the coarse graining transformations to map translational invariant states on $ \mathcal{L}_{\tau}$ into translational invariant states on  $\mathcal{L}_{\tau+1}$ by choosing the elementary isometries of the layer $\tau$ to be all equal. This is represented by the color scheme adopted in Fig. \ref{fig:ttn}, where all the isometries belonging to the same layer have the same color.

\subsection{Computing the reduced density matrix}
\label{subsec rhoA}

Here we briefly describe how to  compute the spectrum of the reduced density matrix $\rho_A$  of a subsystem $A$ 
in a given spin chain \cite{tev-09,atc-11}.
To this aim we should first notice that even when the state $|\Psi\rangle$
is translational invariant, by construction the TTN does not enjoy this symmetry.
As a consequence, there are  choices of the subsystem $A$ for which the computation of  the spectrum of $\rho_A$ is simpler. 
This happens when  $A$ is chosen to be one block (or several  blocks $A_i$) of spins  of the original lattice which are 
coarse grained by the TTN to a single spin at some level $\tilde{\tau} < N$.
For example, given the structure in Fig. \ref{fig:ttn}, it is easier to compute the spectrum of $\rho_A$ made 
by two spins when they belong to the same isometry, since they are coarse grained to a single spin in one layer.
This is due to the fact that, in order to compute the spectrum of $\rho_A$, we do not need to build the full matrix $\rho_A$, but only a simplified matrix $\tilde{\rho}_A$ which is related to $\rho_A$ by similarity transformation, i.e.
\begin{equation}
\rho_A \equiv \Omega_A \, \tilde{\rho}_A \, \Omega_A^{\dagger}\,. 
 \label{eq:t_r}
\end{equation}
The matrix $\Omega_A$ corresponds to the contraction of all the isometries  $\omega$ involved in this coarse-graining process and it transforms each original block $A_i$ into a single spin (see Fig. \ref{fig:rhos}).
The matrices $\tilde{\rho}_A$ are $\chi_{\tilde{\tau}}^m \times \chi_{\tilde{\tau}}^m$, being $\chi_{\tilde{\tau}}$ 
the dimensions of the open bonds in $\tilde{\rho}_A$ and $m$ is the numbers of disjoint intervals composing $A$ 
(in this paper we  will limit have to $m=1$ and $m=2$). 
Although $ \tilde{\rho}_A $ and $\rho_A$ have different dimensions (with $ \tilde{\rho}_A $  smaller than $\rho_A$),  
they have the same spectrum, because they are related by isometries.

\begin{figure}
\vspace{.4cm}
\begin{center}
\includegraphics[width=0.9\textwidth]{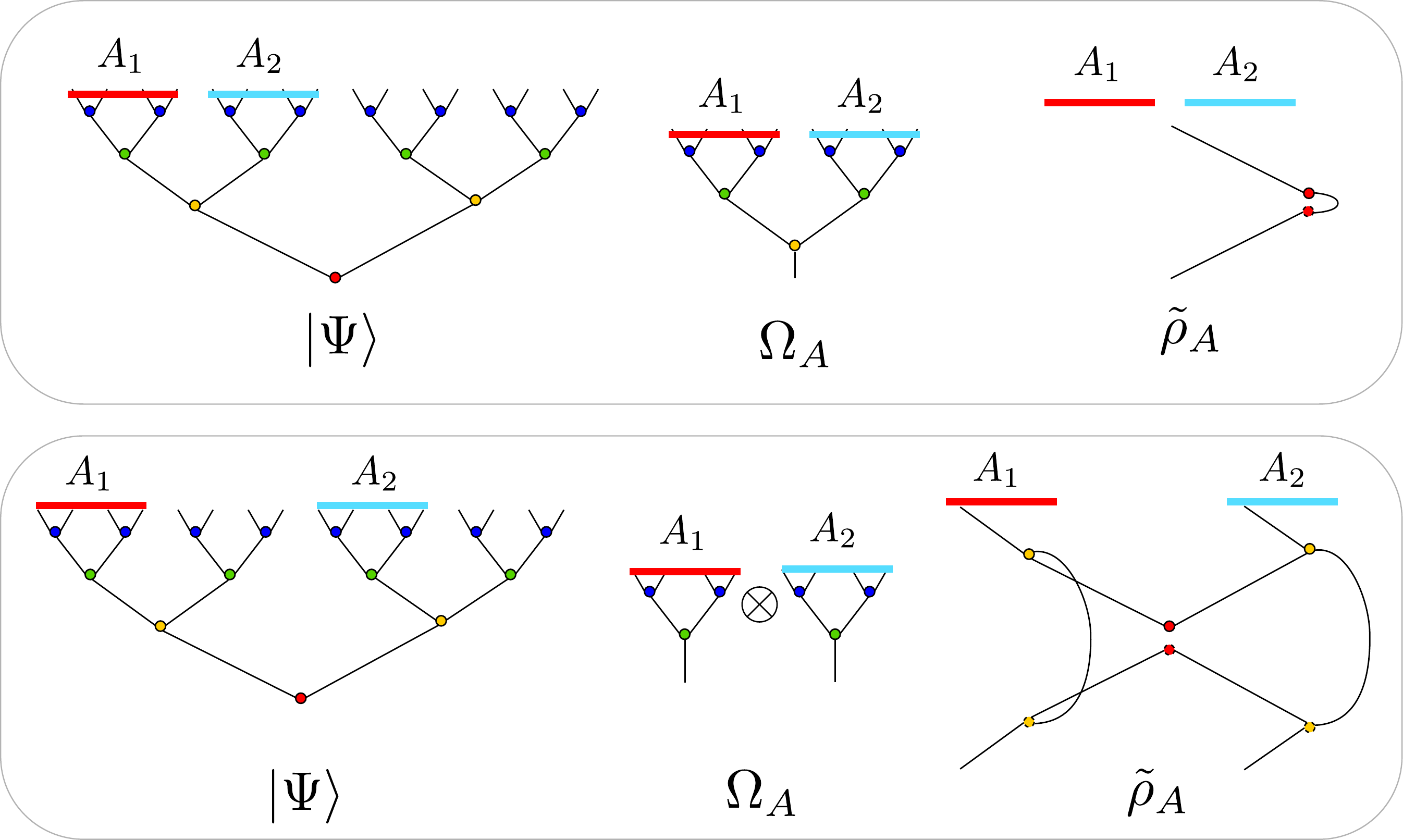}
\end{center}
\caption{Quantities involved in the computation of the $\rho_A$ through a TTN algorithm. Top: $A$ is made by two adjacent intervals.
Bottom: $A$ is made by two disjoint intervals. 
\label{fig:rhos}}  
\end{figure}
 
 Thus, we can focus on the computation of $ \tilde{\rho}_A $. 
In this process, we can identify three groups of tensors: 
(i) the tensors whose support is fully contained in  $B$;
(ii) the tensors whose support is fully contained in $A$; 
(iii) the tensors whose support is shared between $A$ and its complement.
The tensors (i) disappear from the calculation because they are contracted through Eq. (\ref{eq:isometry}).
The tensors (ii) just define $\Omega_A$ and therefore they do not occur in the computation of $ \tilde{\rho}_A $. 
Thus, only the tensors of type (iii) are needed for the computation of $\tilde{\rho}_A$ which is achieved by  
contracting these tensors as shown in Fig. \ref{fig:rhos}.
Once the simplified tensor network encoding $\tilde{\rho_A}$ has been contracted, its diagonalization provides the spectrum 
of  $\rho_A$. The upper bound of the computational cost of the whole algorithm is $\chi_{max}^6$.

Different block configurations (e.g. not optimal choices of $A_1$ and $A_2$ or odd $A_i$ made by odd number of spins) 
can also be obtained from the TTN. 
Naively, one could think that this requires a higher computational cost, but, with some efforts,  it is possible to
envisage more complicated computational strategies to reduce the cost to $\chi_{max}^6$, 
as discussed in Ref. \cite{evenblyvidal11} in a different context.

\subsection{Computing the partial transpose}

\begin{figure}
\vspace{.4cm}
\begin{center}
\includegraphics[width=0.9\textwidth]{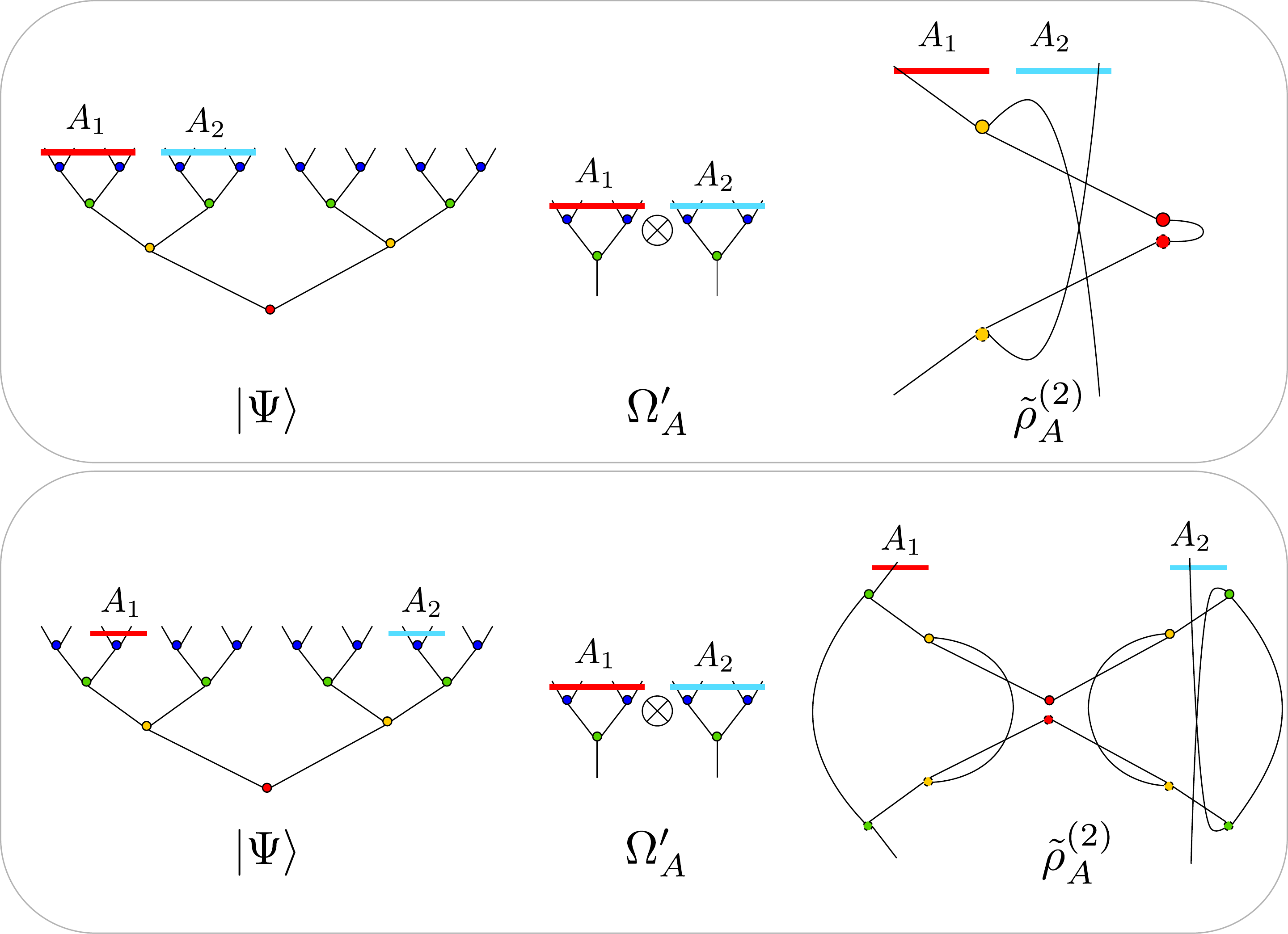}
\end{center}
\caption{Quantities involved in the computation of the $\rho_A^{T_2} $ through a TTN algorithm. Top: $A$ is made by two adjacent intervals.
Bottom: $A$ is made by two disjoint intervals. 
\label{fig:pt_ttn}}
\end{figure}

The calculation of the spectrum of the  partial transpose of $\rho^{T_2}_A$ is very similar to the one explained above for the spectrum of $\rho_A$. 
First we have to choose $A$ as above, namely such that each $A_i$ becomes a single spin at level $\tilde{\tau}$.
Then, we can identify three groups of tensors : (I) the tensors fully contained into $B$; (II) the tensors fully contained into either $A_1$ or $A_2$ (not both); (III) the tensors connecting $A_1$ and $A_2$ or $A_i$ and $B$. 
Notice that, while (I) is the same as (i) above, the group (II) is different from (ii) and, consequently, also (III) is not (iii).
Let us focus on computing
\begin{equation}
\rho_A^{T_2}   \equiv {\Omega'}_A\, \tilde\rho^{(2)}_A\, {\Omega'}_A^{\dagger},
 \label{eq:t_r_t}
\end{equation}
where ${\Omega'}_A$ is an isometric tensor (i.e. ${\Omega'}_A {\Omega'}_A^\dagger = \textrm{id}_{\chi_{\tilde{\tau}}^m}$). 
For the same reasons discussed in the subsection \ref{subsec rhoA}, only the tensor of type (III) enter in the definition of 
$ \tilde\rho^{(2)}_A$.

When  $A$ is made by two disjoint blocks (see Fig. \ref{fig:pt_ttn}, bottom panel), the matrix 
$\tilde{\rho}^{(2)}_A $ is given by  $\tilde{\rho}^{(2)}_A=\tilde{\rho}_A^{T_2}$. 
For two adjacent blocks (see Fig. \ref{fig:pt_ttn}, top panel) this relation does not hold 
because $\tilde{\rho}_A$ has only one incoming and one outgoing bond and 
therefore we cannot distinguish between $A_1$ and $A_2$, which is necessary in order to perform the partial transposition. 
In this case, indeed we have ${\Omega'}_A\neq {\Omega}_A$ and the former reads
\begin{equation}
{\Omega'}_A =\Omega_{A_1} \otimes \Omega_{A_2}.
\end{equation}
Comparing the top panel of Fig. \ref{fig:pt_ttn} (center) to the top panel of Fig. \ref{fig:rhos} (center), we realize that  
the yellow isometry connecting $A_1$ to $A_2$ does not occur in ${\Omega'}_A$, in contrast with ${\Omega}_A$. 
Indeed, the yellow isometry appears in $\tilde{\rho}^{(2)}_A $.
We finally mention that the computational cost of all the algorithm to determine the spectrum of $\rho_A^{T_2}$
is upper-bounded by $\chi_{max}^6$. 


\section{Numerical results for the critical Ising chain in a transverse magnetic field}
\label{Sec5}

In this section we report numerical  results for the transverse field Ising chain obtained 
by means of the  tree tensor network as explained in the previous section. 
We consider chains of finite length $L$ of the form $L=2^M$ and the maximum  
size studied is $L=512$. The $\chi_{max}$ for this simulation has been fixed to $128$, guaranteeing a
relative  precision on the ground state energy of about $10^{-8}$. 
We recall that with the TTN method, using a binary tree, we can quickly access only subsystems of size 
$\ell=2^m$ with $m<M$  integer, as it should be clear from the previous section.

As a first calculation we considered a bipartite chain (i.e. $B\to\emptyset$) and we checked that 
the entanglement negativity reproduces the R\'enyi entropy $S^{(1/2)}_{A_2}$, as it should. 
We do not report explicit plots of these results, but they have been important numerical checks for the 
numerics. 
The results for tripartite systems are reported and discussed in the following subsections.

\subsection{Two adjacent intervals.}

We first consider the case of two adjacent intervals both of equal length $\ell$
in a periodic chain of total length $L$ so that all the 
results depend on the single parameter $z\equiv \ell/L\in[0,1/2]$.
In terms of $z$ and fixing $c=1/2$, the CFT predictions in Eq. (\ref{3ptfinL}) can be written as 
\be\fl
\Tr(\rho_A^{T_2})^n=d_n\times \left\{
\begin{array}{ll}
(L/\pi \sin(\pi z))^{-(n_e/2-2/n_e)/6} 
(L/\pi \sin (2\pi z))^{-(n_e/2+1/n_e)/12}\,,
\\ \\
((L/\pi)^3\sin^2(\pi z) \sin (2\pi z))^{-(n_o-1/n_o)/24},
\end{array}
\right.
\ee
while for the logarithmic negativity  we have from Eq. (\ref{neg3finL})
\be
{\cal E}= \frac18 \ln\Big[\frac{L}\pi \tan(\pi z)\Big] +{\rm cnst} \,.
\ee

\begin{figure}[t]
\begin{center}
\includegraphics[width=.49\textwidth]{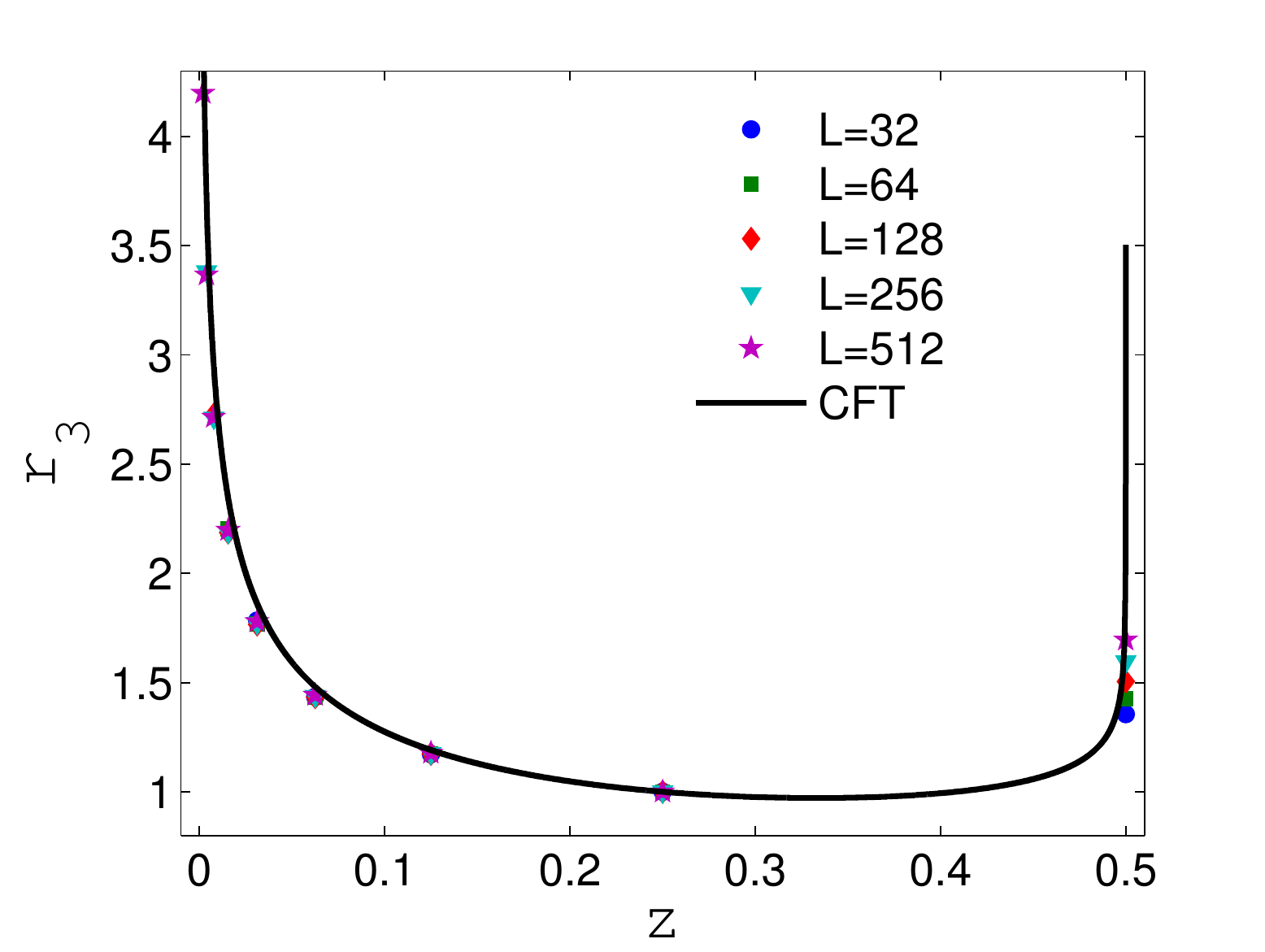}
\includegraphics[width=.49\textwidth]{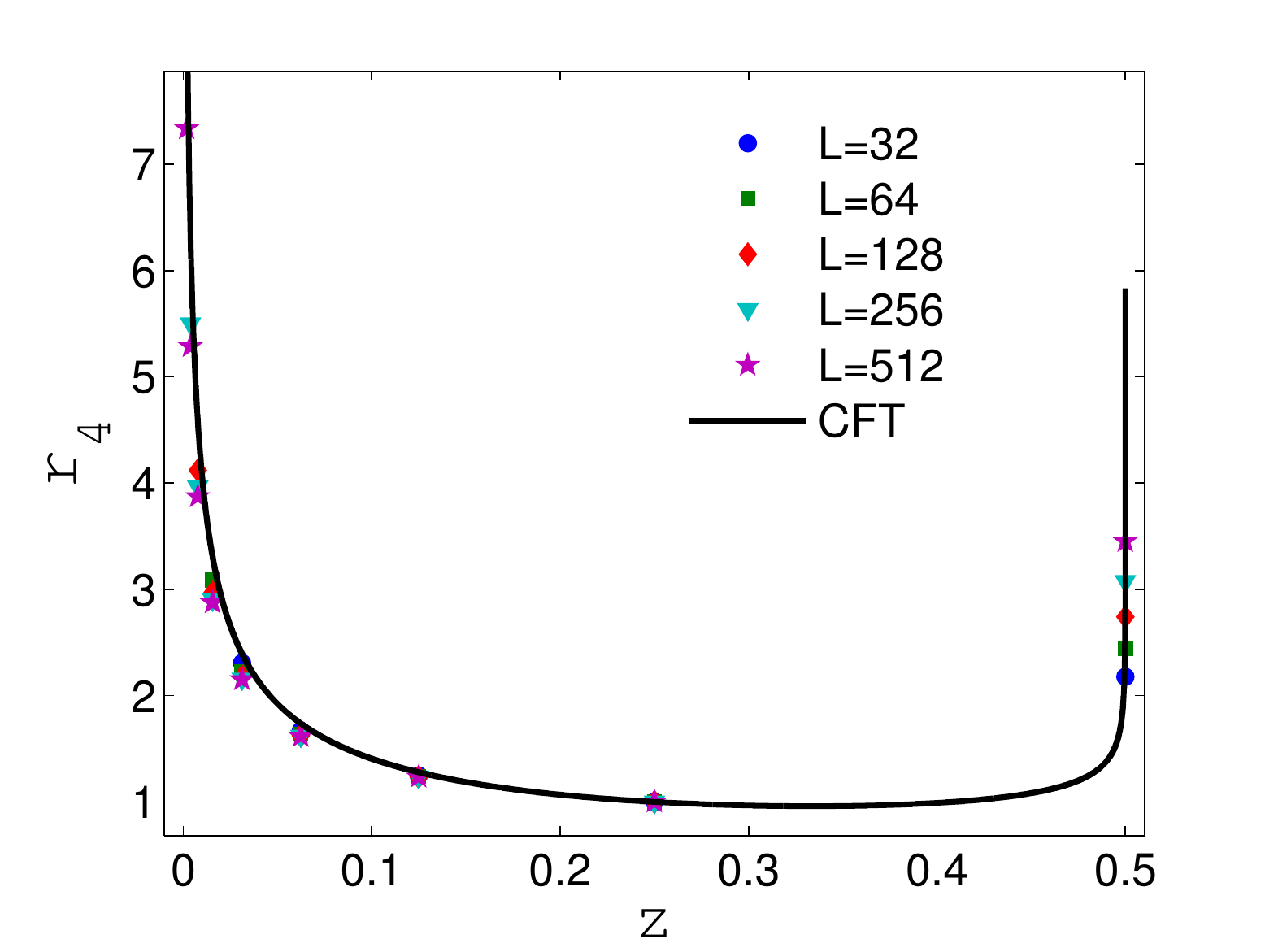}
\end{center}
\caption{
Entanglement for two adjacent intervals of equal length $\ell\leq L/2$ in a periodic chain of length $L$. 
The quantity  $r_n(z)$ in Eq. (\ref{ratiorn})  as function of $z=\ell/L$
compared with the parameter free CFT prediction for $n=3$ (left) and $n=4$ (right).
}
\label{3ptfigr}
\end{figure}

Following Ref. \cite{us-long}, we can construct quantities  in which the dependence on the 
non-universal parameters $d_n$ 
and also the universal dependence on $L$ cancel. 
To this aim, it is enough to divide $\Tr(\rho_A^{T_2})^n$ by the value it assumes 
at a given fixed $\ell$, e.g. $\ell=L/4$, i.e. by considering the quantities
\be
r_n(z)= \ln \frac{\Tr(\rho_A^{T_{A_2=\ell}})^n}{\Tr(\rho_A^{T_{A_2=L/4}})^n},
\label{ratiorn}
\ee
whose parameter free CFT predictions for $n$ even and odd are
\bea
r_{n_e}=\frac1{12}\Big(\frac2{n_e}-\frac{n_e}2  \Big)\ln (2 \sin^2(\pi z))-
\frac1{12} \Big(\frac{n_e}2 +\frac1{n_e} \Big) \ln ( \sin (2\pi z)),\nonumber
\\
r_{n_o}=
\frac1{24}\Big(\frac1{n_o}- n_o\Big)\ln (2\sin^2(\pi z) \sin (2\pi z)).
\label{rnCFT}
\eea
The numerical results for these quantities are shown in Fig. \ref{3ptfigr} for $n=3$ and $n=4$. 
The agreement between the numerical data and the CFT predictions is perfect for all considered values of 
$L$, showing that finite size corrections are very small for these quantities. 
Notice that for $z=1/2$ we have a bipartite system (i.e. $B\to \emptyset$) and the equations in 
(\ref{rnCFT}) obviously do not work since the data are described by Eqs. (\ref{cftTr1int2}) and (\ref{cftTr1int}),
reflecting the fact that the limit $z\to1/2$ is approached in a non-uniform way (i.e. the limits $z\to 1/2$ and 
$N\to\infty$ do not commute as obvious). 

For the logarithmic negativity, we can analogously define the subtracted quantity
\be
\e(z)={\cal E}(\ell,L)-{\cal E}(L/4,L)=\frac18 \ln[ \tan(\pi z)] \,,
\label{epsdef}
\ee
and again the r.h.s. is a parameter free CFT prediction.
In Fig. \ref{3ptfigeps}, this prediction is compared with the numerical data 
and the agreement is extremely good except at $z=1/2$ where the numerical data
are described by the bipartite formula (\ref{EN1int}). 
Notice that finite size scaling corrections are even smaller than those for the 
quantities $r_n(z)$ in Fig. \ref{3ptfigr}.

\begin{figure}[t]
\center{\includegraphics[width=.69\textwidth]{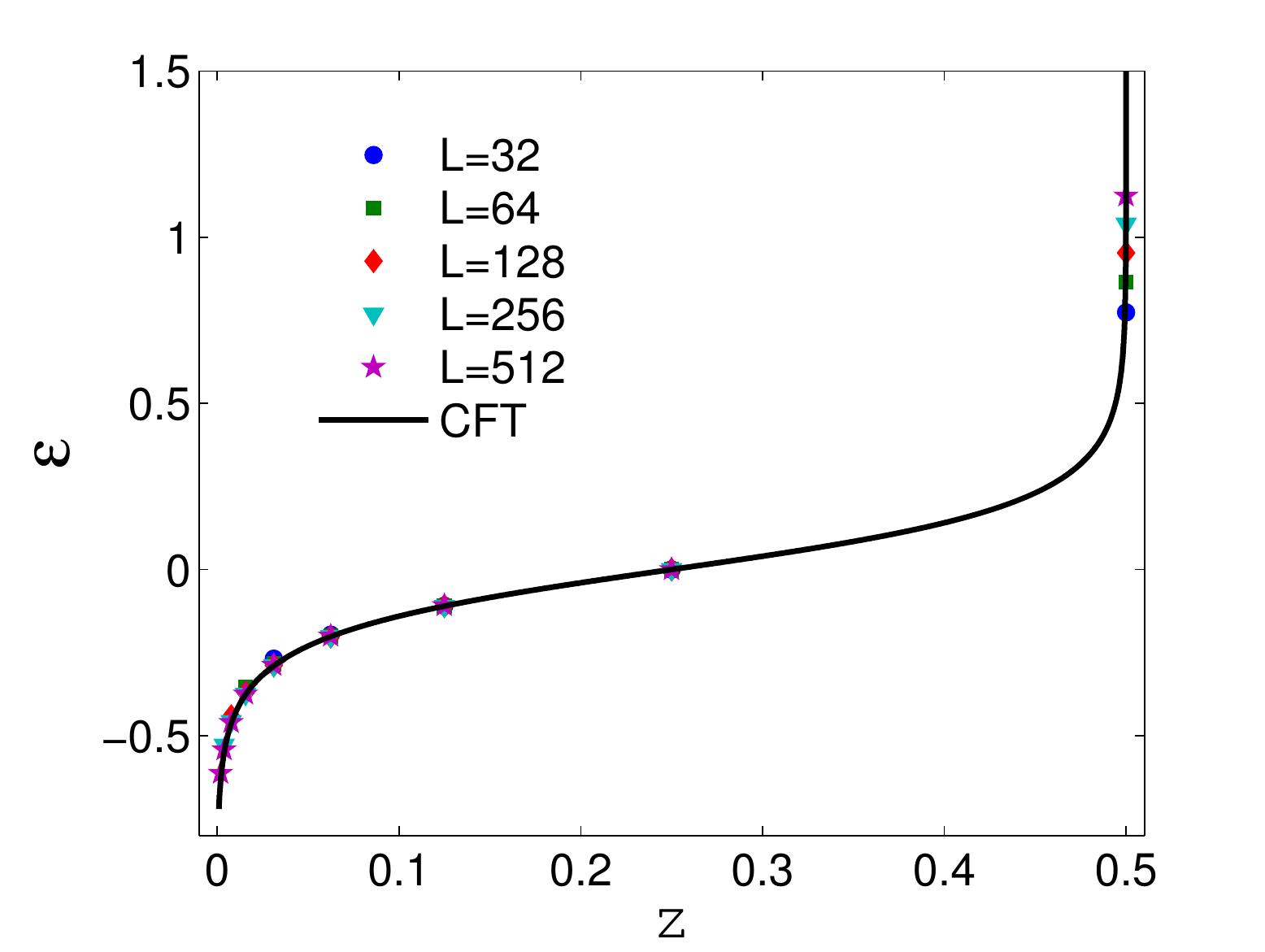}}
\caption{Entanglement negativity for two adjacent intervals of equal length $\ell<L/2$ in a periodic chain of length $L$: 
Subtracted negativity $\e(z)$ in Eq. (\ref{epsdef}) compared with the parameter free CFT prediction.
}
\label{3ptfigeps}
\end{figure}

\subsection{Two disjoint intervals}

In this section we study the most interesting and difficult situation of 
two disjoint intervals for which an accurate numerical study of the negativity has been already performed 
by means of density matrix renormalization group in Ref. \cite{Neg1}, 
but before the systematic CFT derivation in Refs. \cite{us-letter,us-long}.
Here we first consider the traces $\Tr(\rho_A^{T_2})^n$ and $\Tr(\rho_A)^n$.
Indeed, although standard R\'enyi entropies given by $\Tr \rho_A^n$ have been already 
studied in Refs. \cite{cct-11,atc-10,fc-10}, they show large corrections to the scaling which is
worth recalling before embarking in the analysis of $\Tr(\rho_A^{T_2})^n$.

In order to determine numerically the function ${\cal F}_n(x)$, we calculate for several finite chains the quantity 
\be
F^{\rm lat}_n(x)= \frac{\Tr \rho_{A_1\cup A_2}^n}{\Tr \rho_{A_1}^n \Tr \rho_{A_2}^n} (1-x)^{(n-1/n)/12 }\,,
\label{Flat}
\ee
which in the scaling limit should converge to the CFT prediction ${\cal F}_n(x)$.
In a finite system of length $L$, the four-point ratio $x$ must be rewritten by replacing all distances  by the corresponding 
chordal lengths. For two intervals of the same length $\ell$ at distance $r$ this reads
\be
x=\left(\frac{\sin(\pi\ell/L)}{\sin(\pi (\ell+r)/L)}\right)^2\,.
\label{xFS}
\ee

\begin{figure}[t]
\begin{center}
\includegraphics[width=.49\textwidth]{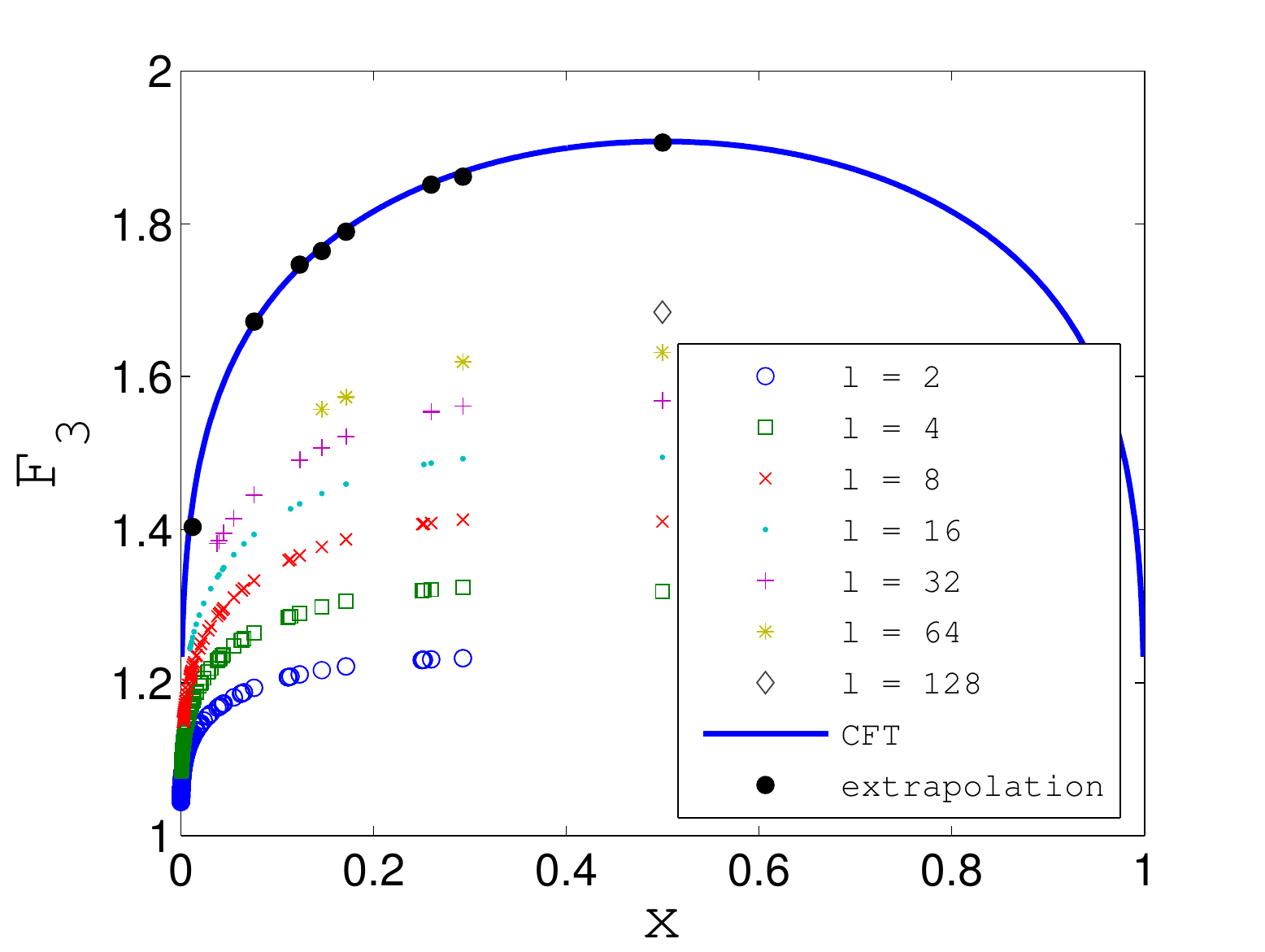}
\includegraphics[width=.49\textwidth]{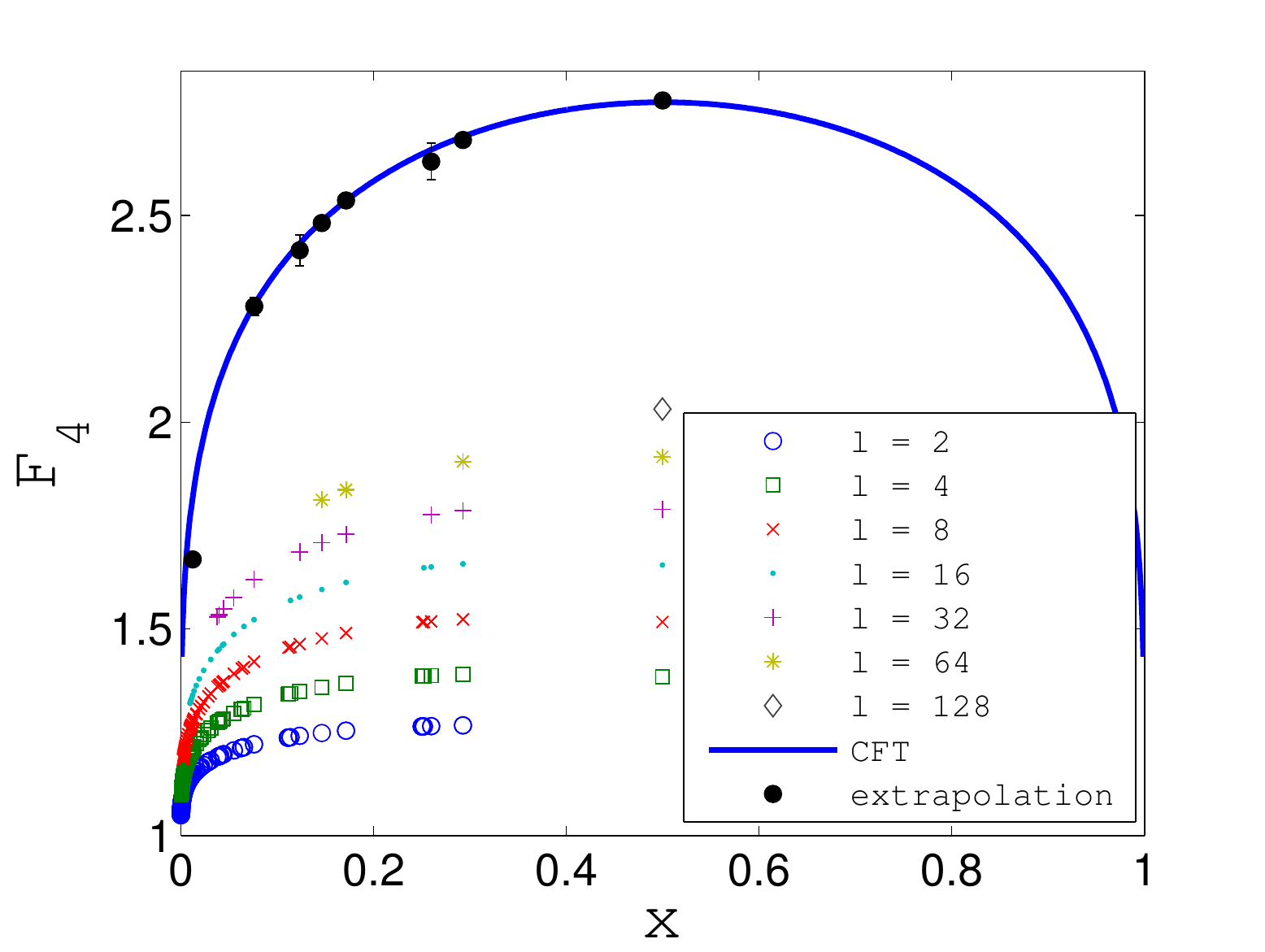}
\center{\includegraphics[width=.49\textwidth]{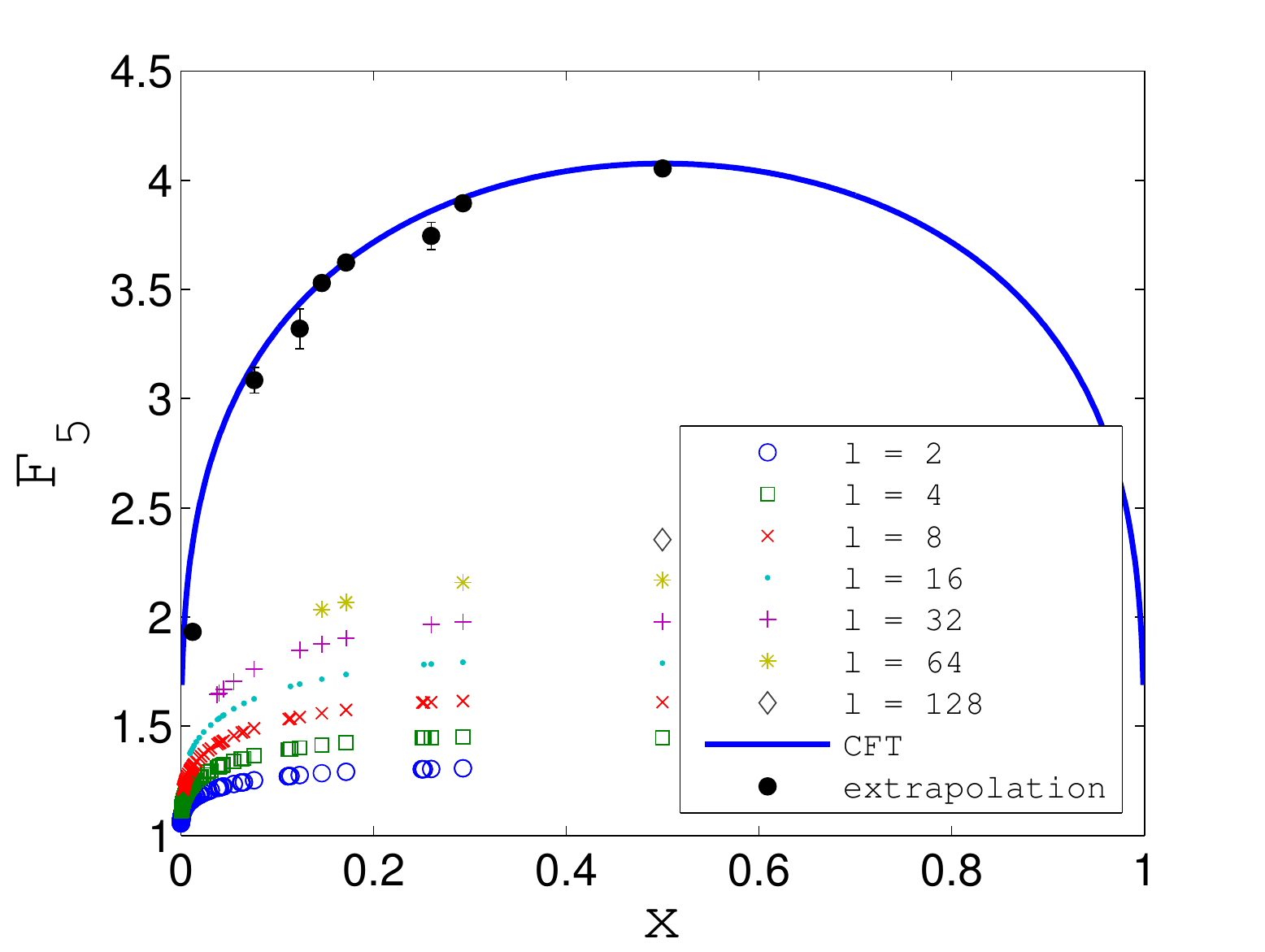}}
\end{center}
\caption{Numerical results for $F_n^{\rm lat}(x)$ as function of $x$ for different values of $\ell$ for $n=3$ (top left),
 $n=4$ (top right), and $n=5$ (bottom).
The data are extrapolated to $\ell\to\infty$ by means of Eq. (\ref{ansatz2}).
The extrapolated data (topmost set of data) are in excellent agreement with the CFT prediction (continuous
line). 
}
\label{F3F4}
\end{figure}

The numerical data for the function $F^{\rm lat}_n(x)$ are reported in Fig. \ref{F3F4} for $n=3,4,5$ 
as function of $x$ for various values of $\ell$ (i.e. different values of $L$ according to Eq. (\ref{xFS})). 
It is evident that strong scaling corrections affect the data, as known  from previous analyses 
\cite{atc-10,fc-10}.
It has been argued  on the basis of the general CFT arguments \cite{cc-10}, 
and shown explicitly in few examples \cite{ccen-10,ce-10,un-vari} both analytically and numerically, that the entanglement entropies 
(hence also the function $F^{\rm lat}_n(x)$), display `unusual' corrections to the scaling which, at the leading order, 
can be effectively taken into account by the scaling ansatz 
\begin{eqnarray}
F_n^{\rm lat}(x)={\cal F}_n(x)+\ell^{-1/n}f_n(x)+\dots\,.
\label{ansatz}
\end{eqnarray}
However, the corrections to the scaling in Fig. \ref{F3F4} cannot be captured by this simple ansatz
because subleading corrections to the scaling become more and more important with increasing the 
value of $n$. 
Indeed, corrections of the form  $\ell^{-m/n}$  for any integer $m$ are know to 
be present \cite{ce-10,fc-10,atc-11}. 
Thus  the most general finite-$\ell$ ansatz is
\be
F_n^{\rm lat}(x)={\cal F}_n(x)+\frac{f_n^{(1)}(x)}{\ell^{1/n}} +\frac{f_n^{(2)}(x)}{\ell^{2/n}}+\frac{f_n^{(3)}(x)}{\ell^{3/n}}\dots\,.
\label{ansatz2}
\ee
The effect of the subleading corrections is enhanced by the fact the amplitude functions $f_n^{(i)}(x)$ 
have different signs determining a non-monotonic behavior in $\ell$ (in particular, we have that $f^{(1)}_n$ is always negative, 
while $f^{(2)}_n$ and $f^{(3)}_n$ are always positive, as discussed in Ref. \cite{atc-11}).
In order to have an accurate extrapolation to $\ell\to \infty$, for any $n$ we consider all the 
corrections above up to order $O(\ell^{-3/n})$ and 
we get the extrapolations reported in Fig. \ref{F3F4}.
The error bars are estimated by studying the stability of the extrapolation with respect to the number of sizes $\ell$ included in the fit. 
The overall agreement of the extrapolated points with the CFT prediction is excellent for all values of $x$ and for the 
three considered values of $n$, reproducing the results in Refs. \cite{atc-10,fc-10}.

\begin{figure}[t]
\begin{center}
\includegraphics[width=.49\textwidth]{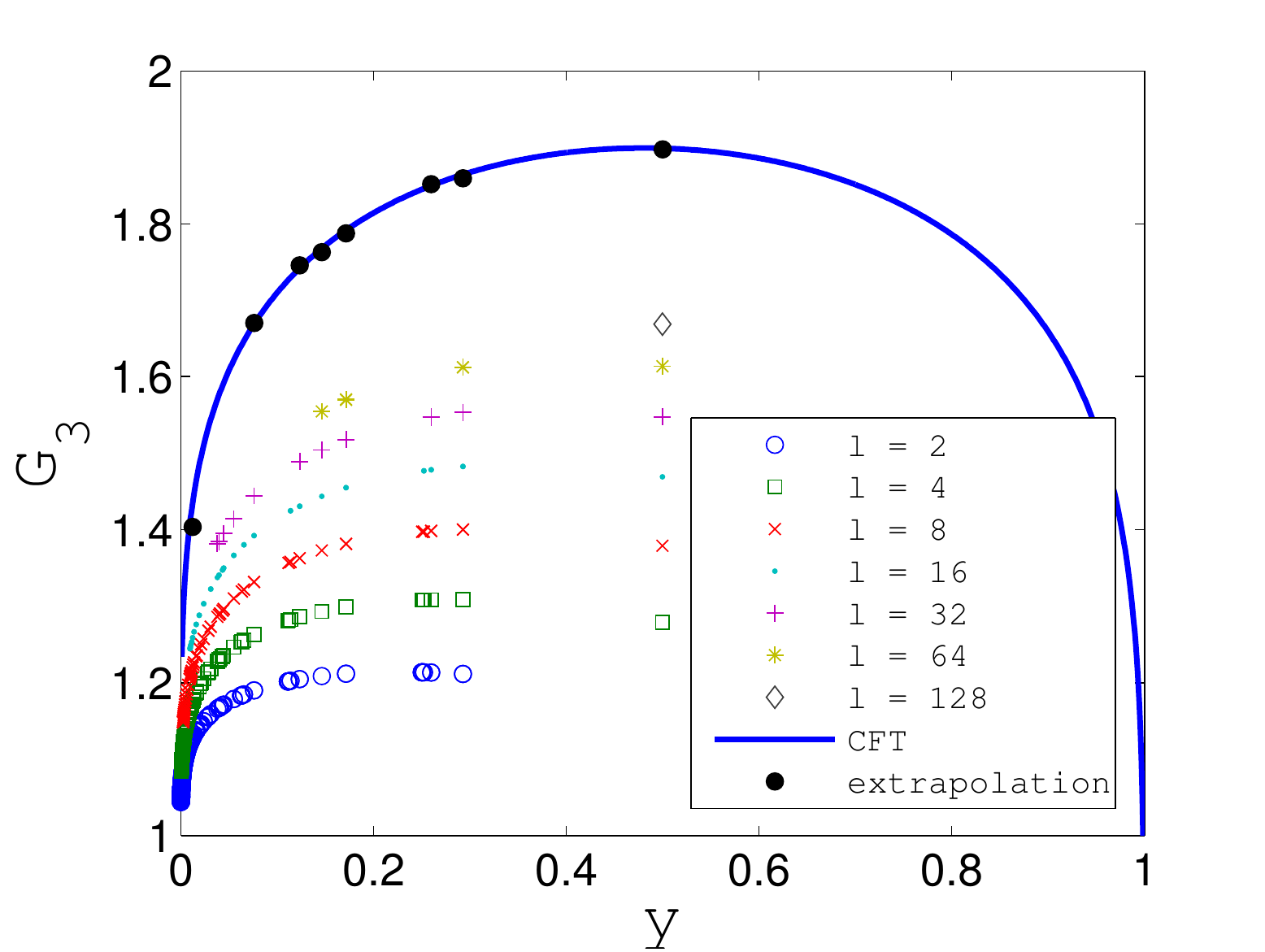}
\includegraphics[width=.49\textwidth]{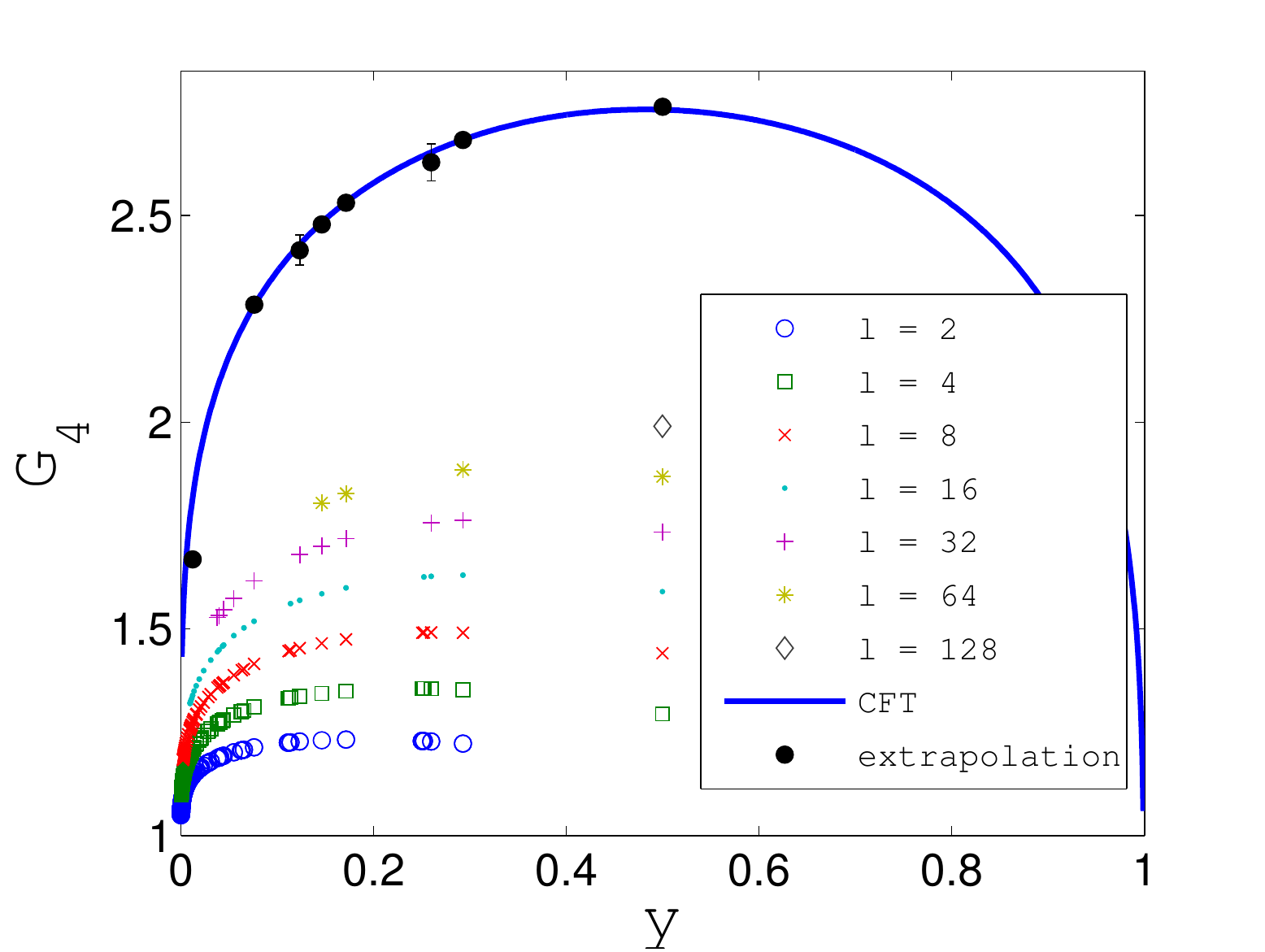}
\center{\includegraphics[width=.49\textwidth]{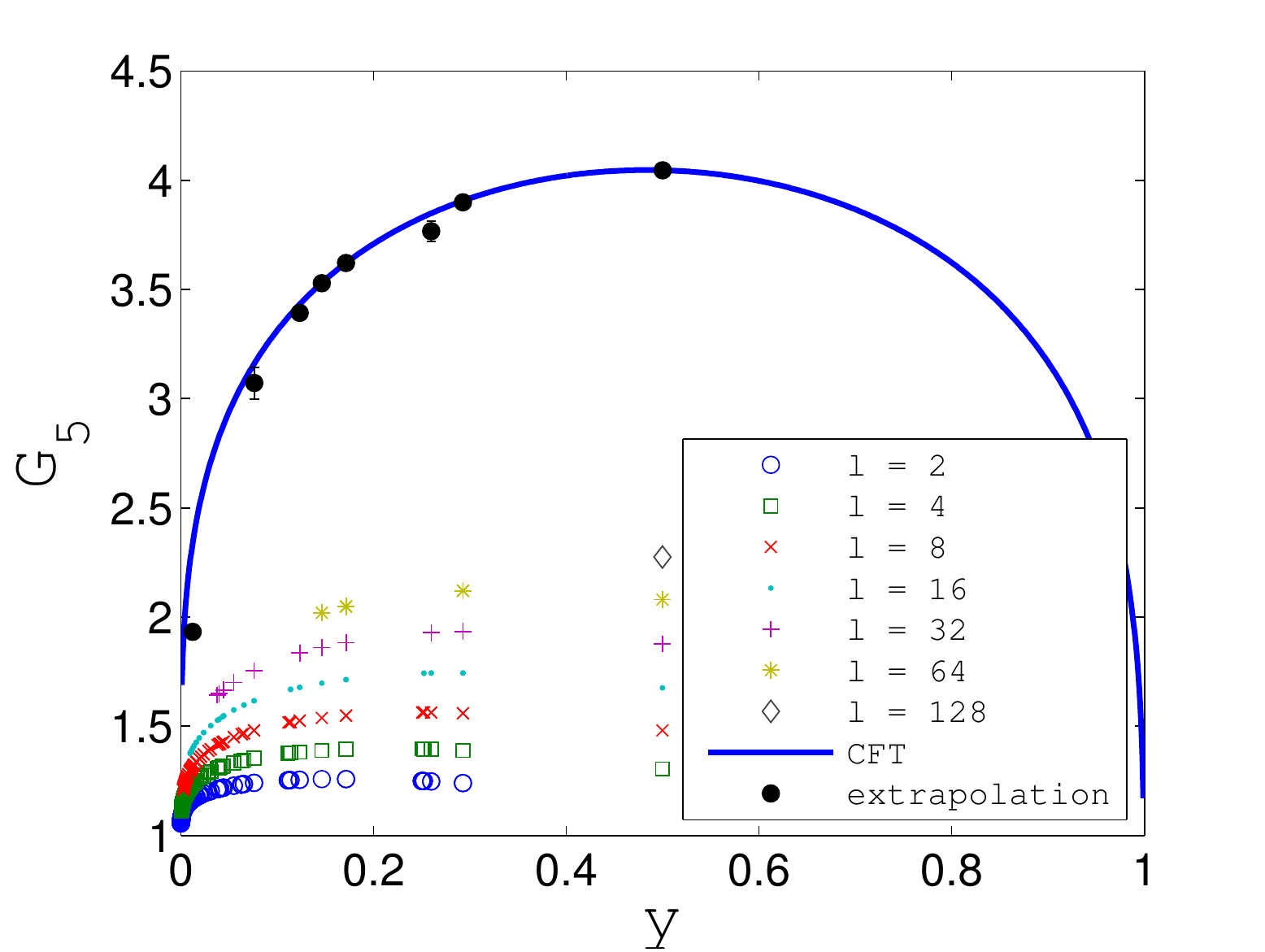}}
\end{center}
\caption{Numerical results for $G_n^{\rm lat}(x)$ as function of $y$ for different values of $\ell$ for $n=3$ (top left),
 $n=4$ (top right), and $n=5$ (bottom).
The data are extrapolated to $\ell\to\infty$ by means of Eq. (\ref{ansatz2G}).
The extrapolated data (topmost set of data) are in excellent agreement with the CFT prediction (continuous
line).
}
\label{G3G4}
\end{figure}

After having summarized the corrections to the scaling for the entanglement entropies we can 
turn to the integer powers of the partial transpose in which we are interested here. 
In analogy with Eq. (\ref{Flat}) we can define the lattice ratio 
\be
G^{\rm lat}_n(y)= \frac{\Tr (\rho_{A_1\cup A_2}^{T_2})^n}{\Tr \rho_{A_1}^n \Tr \rho_{A_2}^n} (1-y)^{(n-1/n)/12 }\,,
\label{Glat}
\ee
that in the limit $\ell\to\infty$ is expected to converge to the CFT scaling function 
${\cal G}_n(y)$ given by Eq. (\ref{Gn ising}). 
In the case at hand, the numerical value of $y$ is given by the same expression in Eq. (\ref{xFS}) for $x$.
The numerical data for $G^{\rm lat}_n(y)$ are reported in Fig. \ref{G3G4} for $n=3,4,5$ as function of 
$y$ for various values of $\ell$. 
Also in this case, large scaling corrections are present. 
These are expected to be of the same form as for $F_n^{\rm lat}(x)$, i.e. described by the ansatz
\be
G_n^{\rm lat}(y)={\cal G}_n(y)+\frac{g_n^{(1)}(y)}{\ell^{1/n}} +\frac{g_n^{(2)}(y)}{\ell^{2/n}}+\frac{g_n^{(3)}(y)}{\ell^{3/n}}\dots\,.
\label{ansatz2G}
\ee
We repeat exactly the same analysis as for $F_n^{\rm lat}(x)$ to extrapolate the data to $\ell\to\infty$.
As above, we find that $g^{(1)}_n$ is always negative, while $g^{(2)}_n$ and $g^{(3)}_n$ are positive. 
Using this observation, we extrapolate to $\ell\to\infty$, obtaining  the results (with error bars) reported in 
Fig. \ref{G3G4}. These points agree very well with the 
CFT prediction in Eq. (\ref{Gn ising}) for the three values of $n$ considered.

\begin{figure}[t]
\begin{center}
\includegraphics[width=.49\textwidth]{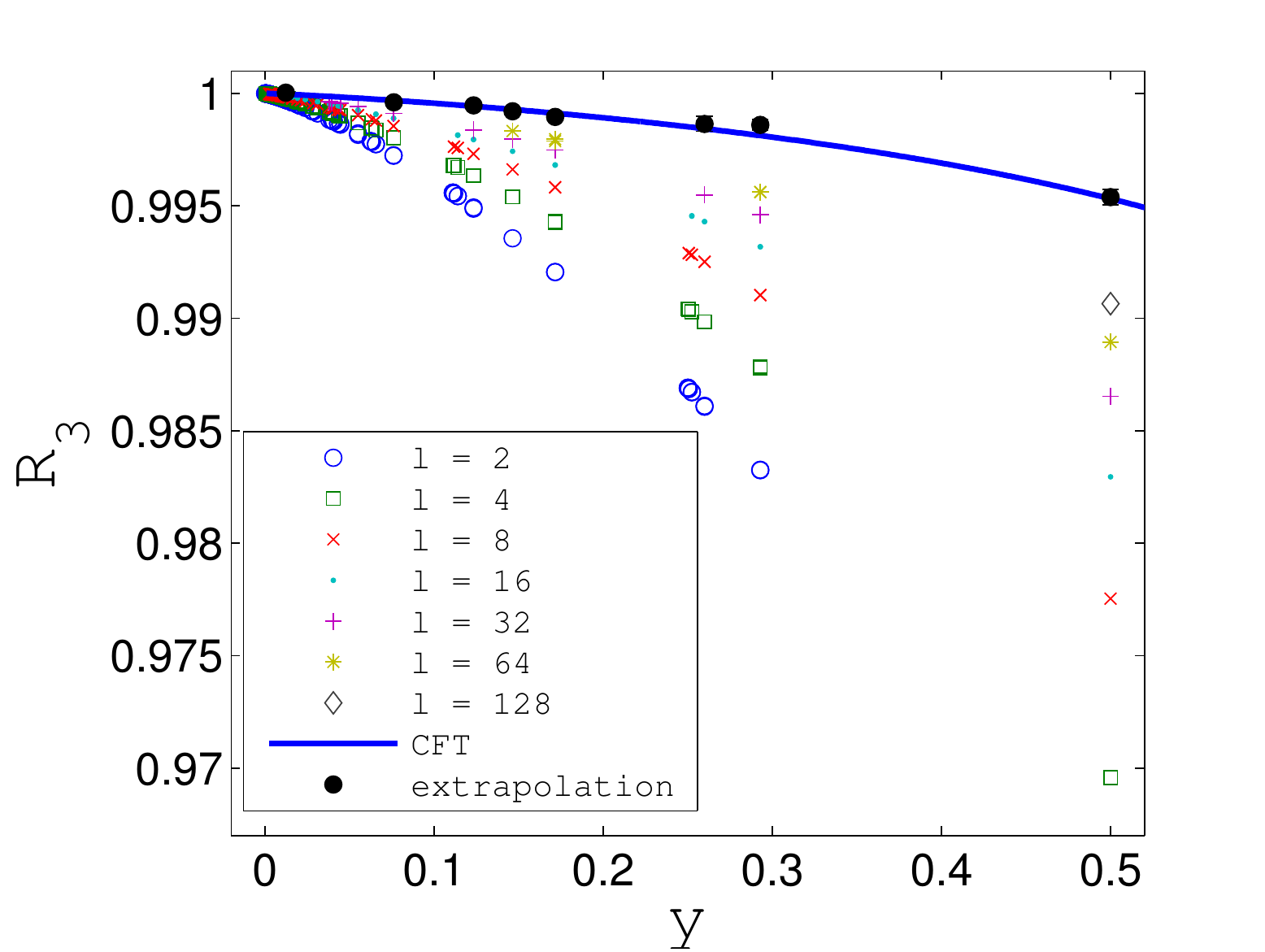}
\includegraphics[width=.49\textwidth]{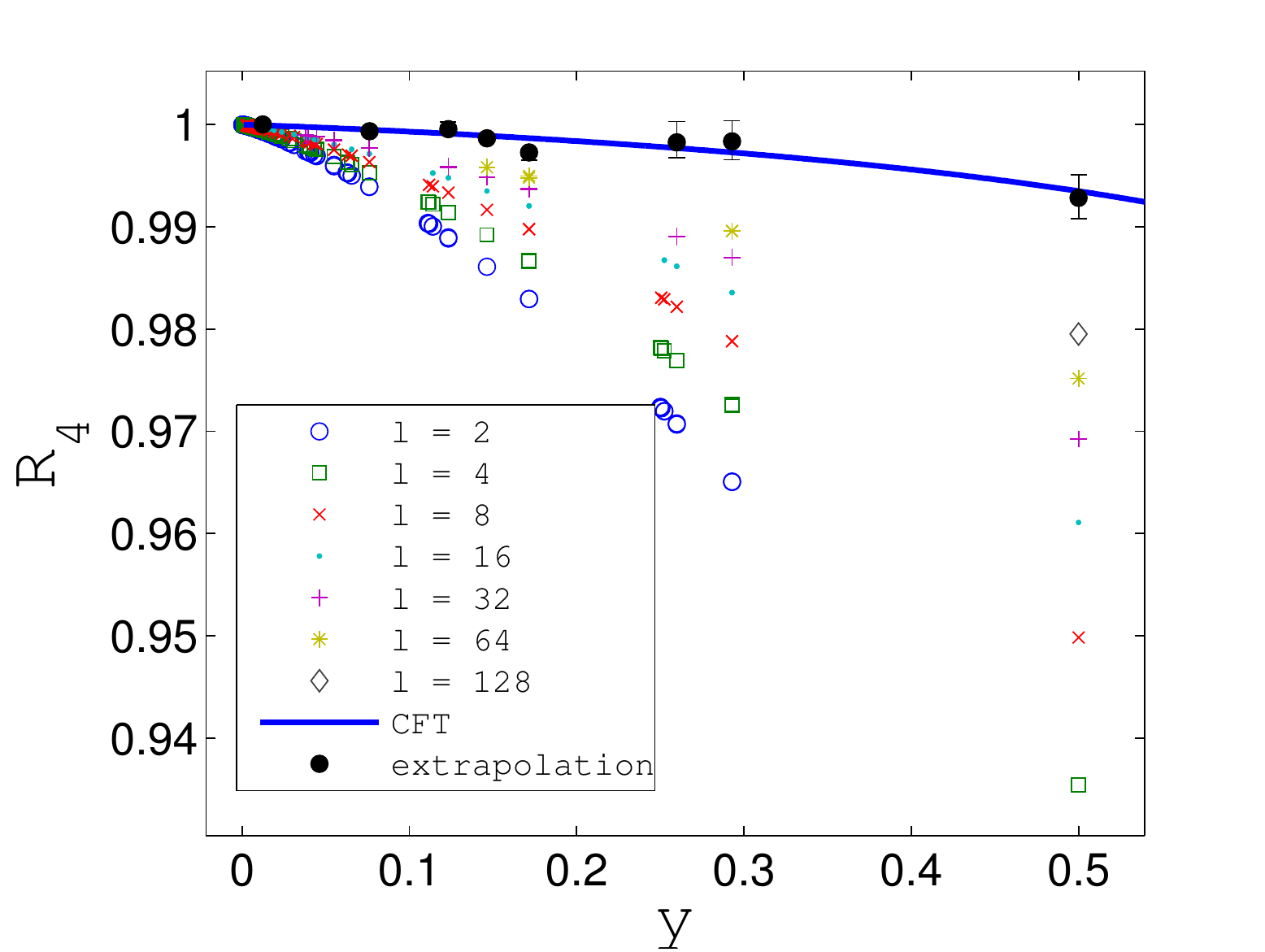}
\center{\includegraphics[width=.49\textwidth]{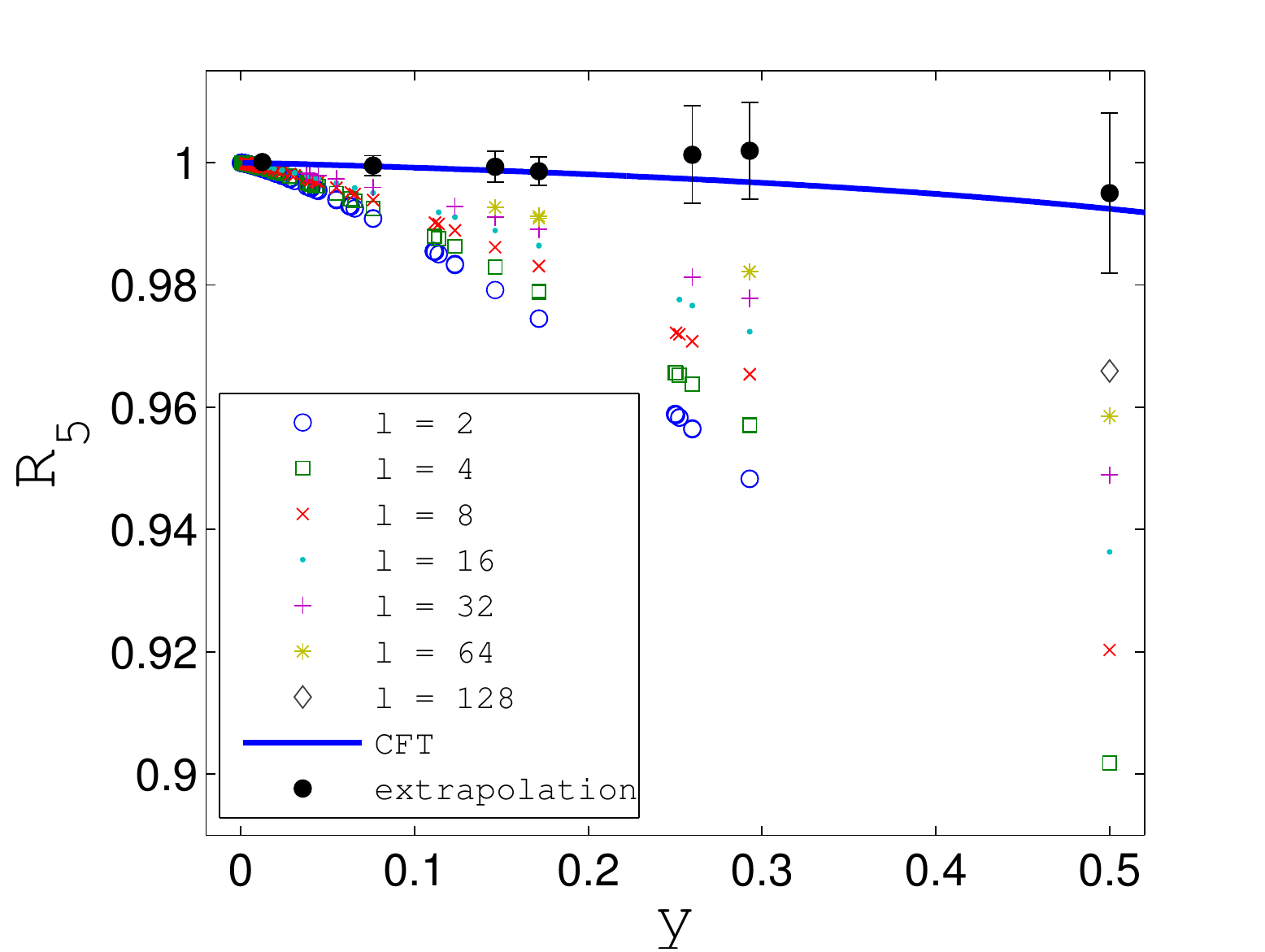}}
\end{center}
\caption{Numerical results for $R_n^{\rm lat}(x)$ as function of $y$ for different values of $\ell$ for $n=3$ (top left),
 $n=4$ (top right), and $n=5$ (bottom).
The data are extrapolated to $\ell\to\infty$ by means of Eq. (\ref{ansatz2R}).
The extrapolated data (topmost set of data) are in excellent agreement with the CFT prediction (continuous
line).
}
\label{R3R4}
\end{figure}

It is evident from the comparison of Figs. \ref{F3F4} and \ref{G3G4} that the functions $G^{\rm lat}_n(y)$ and 
$F^{\rm lat}_n(y)$ are very close to each other and the logarithmic negativity is instead only determined by the small differences 
between the two. 
Thus, as already discussed in Sec.~\ref{Sec3}, a practical way 
to have more accurate tests of the CFT predictions which are sensitive to the small differences between 
the two universal functions ${\cal G}_n(y)$ and ${\cal F}_n(y)$ is to consider the ratio
$R_n(y)$ between the two and the analogous lattice quantity
\be
R_n^{\rm lat}(y)\equiv \frac{G_n^{\rm lat}(y)}{F_n^{\rm lat}(y)},
\ee
which in the limit $\ell\to \infty$ converges to the CFT prediction  in Eq. (\ref{Rn ising def}).
The numerical data for $R^{\rm lat}_n(y)$ are reported in Fig. \ref{R3R4} for $n=3,4,5$ as function of 
$y$ for different values of $\ell$. 
Once again, large scaling corrections are present and there are no accidental cancellations
in the ratio, so that they are again expected to be of the same form as for $F_n^{\rm lat}(x)$, 
i.e. described by the ansatz
\be
R_n^{\rm lat}(y)=R_n(y)+\frac{r_n^{(1)}(y)}{\ell^{1/n}} +\frac{r_n^{(2)}(y)}{\ell^{2/n}}+\frac{r_n^{(3)}(y)}{\ell^{3/n}}\dots\,.
\label{ansatz2R}
\ee
We repeat again the same analysis as for $F_n^{\rm lat}(x)$ to extrapolate the data to $\ell\to\infty$
and the results (with error bars) are reported in Fig. \ref{R3R4}. 
Unlike $f^{(j)}_n(x)$'s and $g^{(j)}_n(y)$'s, in this case the signs of $r^{(j)}_n(y)$'s are not defined (indeed $r^{(j)}_n$'s 
can be written as complicated combinations of $f^{(j)}_n$'s and $g^{(j)}_n$'s).
For this reason, the error bars in Fig. \ref{R3R4} are larger than the ones in Fig. \ref{F3F4} and in Fig. \ref{G3G4}.
It is evident that  the extrapolated points in Fig. \ref{R3R4} agree very well with the CFT prediction  for the three 
considered values of $n$.
It is very remarkable that the numerical calculations are accurate enough to detect 
the small differences of  these ratios from $1$ (at least for $n=3$ and $n=4$, while for 
$n=5$ the estimated error is too large to distinguish the extrapolation from one).

\begin{figure}[t]
\begin{center}
\includegraphics[width=.69\textwidth]{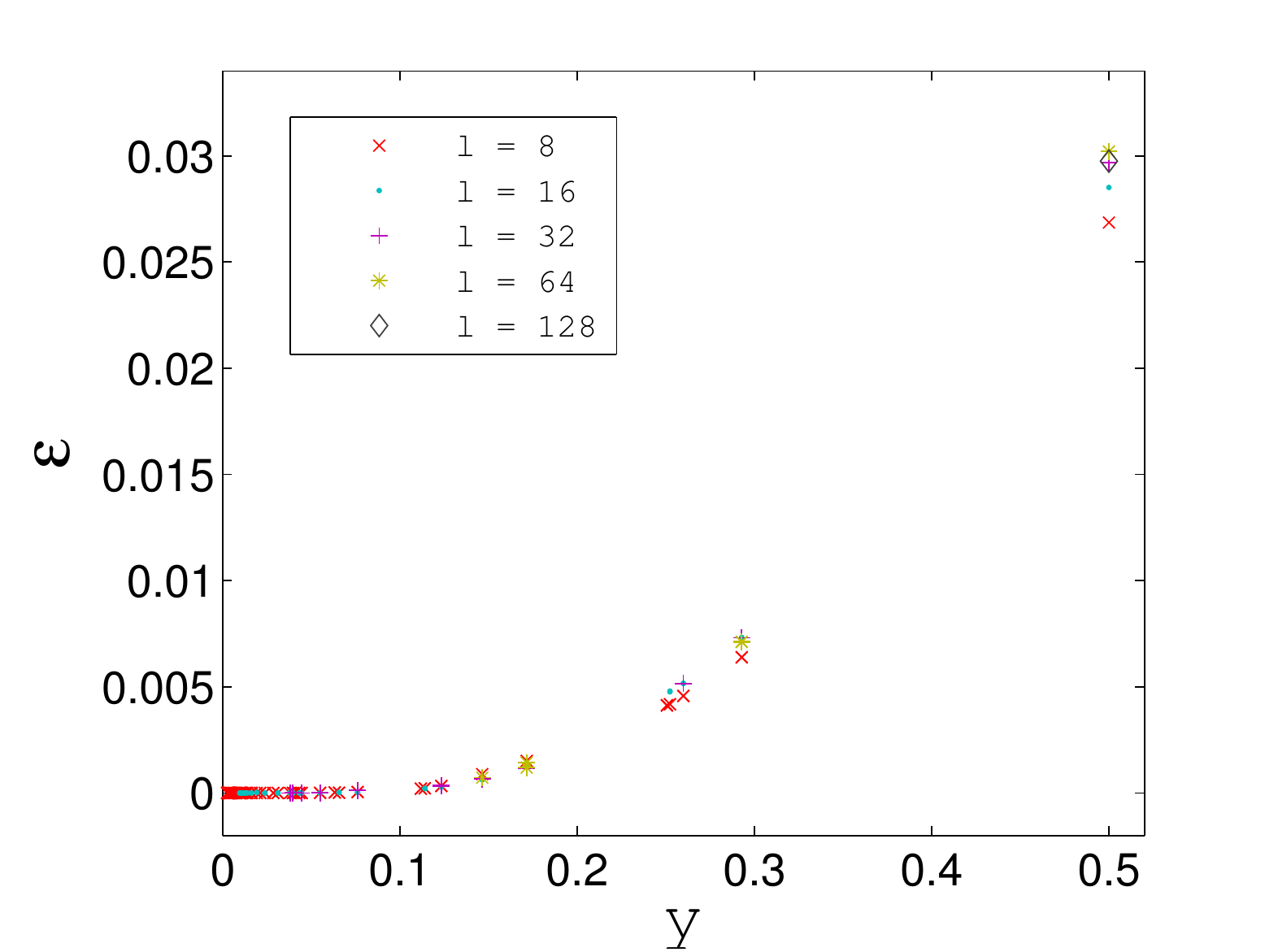}
\end{center}
\caption{Logarithmic negativity for two intervals of equal length $\ell$ at distance $r$ as function 
of the four point ratio $y$. 
}
\label{neg2bl}
\end{figure}

Finally we turn to the study of  the logarithmic negativity ${\cal E}$.
The numerical data  as a function of $y$ are reported in Fig.~\ref{neg2bl} for several values of $\ell$.
In the figure all data collapse on a single curve, with some tiny corrections to the scaling for the smaller
values of $\ell$, which however are much smaller than the scaling corrections for the ratios $R_n(y)$,
as evident from a comparison between Fig.~\ref{neg2bl} and Fig.~\ref{R3R4}.
This is similar to what has been observed for the free boson in Ref. \cite{us-long} and 
recalls the similar effect for the entanglement entropies both for one and two intervals 
where the unusual corrections to the scaling are suppressed for the von Neumann entropy 
for periodic boundary conditions \cite{ccen-10,ce-10}. 
It would be then interesting to study the negativity for Ising chains with open boundary conditions to 
check whether the unusual corrections are enhanced (analogously to Friedel oscillations \cite{lsca-06,fc-11}). 
On the more theoretical side,  we have not been able to find the analytic continuation of $R_{n_e}(y)$
to $n_e\to1$ which would allow a strict check of  the numerical data in Fig. \ref{neg2bl}.  
However, the data for  $y\ll1$ are very close to zero in agreement with the general result \cite{us-long}
that the negativity should fall off faster than any power with the separation of the intervals. 

\section{Conclusions}
\label{Sec6}

In this manuscript we presented a systematic study of the entanglement negativity and 
of the traces of integer powers of the partial transpose of the reduced density matrix in the critical Ising chain.
In order to have an accurate numerical evaluation of these quantities we adapted the tree tensor network technique
to the calculation of the eigenvalues of the partially transposed reduced density matrix. 
For two adjacent intervals we found perfect agreement with the model independent CFT predictions
in Refs. \cite{us-letter, us-long}.
For two disjoint intervals we first derived a CFT prediction for $\Tr (\rho_{A_1\cup A_2}^{T_2})^n$ 
which later has been compared with the numerical data taking into account the finite size corrections 
induced by the finite length of the blocks. 
Unfortunately, in order to determine the logarithmic negativity, it remains a hard open problem to find 
the analytic continuation to $n_e\to1$ of these traces. 
This reflects the similar problem found for the standard entanglement entropies, 
where explicit formulas for $\Tr \rho_{A_1\cup A_2}^n$ have been obtained for some models \cite{cct-09,cct-11}, 
but they have not been analytically continued to $n\to1$, preventing us to write down a close formula for the von Neumann entropy.

\section*{Acknowledgments} 
We are very grateful to John Cardy for many discussions on the subject of the paper. 
ET would like to thank Andrea Coser and Maurizio Fagotti for useful discussions.
We thank Vincenzo Alba for correspondence on a related work \cite{vinc-new}. 
All authors thank the Galileo Galilei Institute in Florence where this work was initiated. 
ET thanks the Dipartimento di Fisica dell'Universit\`a di Pisa, Scuola Normale Superiore di Pisa 
and ICFO in Barcelona for the warm hospitality. LT thanks SISSA in Trieste for hospitality during the last part of this work. 
This work was supported by the 
ERC under  Starting Grant  279391 EDEQS (PC) and by TOQATA (FIS2008-00784),
FP7-PEOPLE-2010-IIF ENGAGES 273524, ERC QUAGATUA, EU AQUTE (LT).


\end{document}